\documentclass[aps,prl,twocolumn,superscriptaddress,floatfix,citeautoscript,preprintnumbers]{revtex4-1}
\usepackage{graphicx,xcolor}
\usepackage{amssymb}
\usepackage{amsmath}
\usepackage{epstopdf}
\usepackage[normalem]{ulem}
\usepackage{comment}
\usepackage{breakcites}
\usepackage[breaklinks=true]{hyperref}
\usepackage[caption=false]{subfig}
\usepackage{enumitem}
\usepackage{soul}
\usepackage{bbold,dsfont}

\def\Z2{Z_{\mathrm{st2}}}

\newcommand{\beq}{\begin{equation}}
\newcommand{\eeq}{\end{equation}}
\newcommand{\bal}{\begin{align}}
\newcommand{\eal}{\end{align}}

\newcommand{\ket}[1]{\mbox{$ | #1 \rangle $}}

\newcommand{\beqa}{\begin{eqnarray}}
\newcommand{\eeqa}{\end{eqnarray}}

\definecolor{MyDarkBlue}{rgb}{0,0.08,0.45} 
\definecolor{MyLightMagenta}{cmyk}{0.1,0.8,0,0.1} 
\definecolor{MLM}{cmyk}{0.1,0.8,0,0.1} 
\definecolor{MyDarkGreen}{rgb}{0,0.45,0.08} 
\definecolor{MDG}{rgb}{0,0.55,0.05} 
\definecolor{lightcarminepink}{rgb}{0.9, 0.4, 0.38}

\newcommand{\figref}[1]{Fig.~\ref{#1}}

\usepackage[textsize=footnotesize]{todonotes}

\definecolor{atomictangerine}{rgb}{1.0, 0.6, 0.4}
\definecolor{bluegray}{rgb}{0.4, 0.6, 0.8}
\definecolor{brightube}{rgb}{0.82, 0.62, 0.91}
\definecolor{brilliantlavender}{rgb}{0.96, 0.73, 1.0}

\graphicspath{../results/}

\begin{document}

\title{Simulating 2+1d $\mathds{Z}_3$ lattice gauge theory with iPEPS} 

\begin{abstract}
	We simulate a zero-temperature pure $\mathds{Z}_3$ Lattice Gauge Theory in 2+1
dimensions by using an iPEPS (Infinite Projected Entangled-Pair State) ansatz
for the ground state. Our results are therefore directly valid in the thermodynamic limit.
They clearly show two distinct phases separated by a phase transition. 
We introduce an update strategy that enables plaquette terms and Gauss-law constraints to be applied as sequences of two-body operators. This allows the use of the most up-to-date iPEPS algorithms.
From the calculation of spatial Wilson loops
 we are able to prove the existence of a confined phase. We show that with relatively low computational
cost it is possible to reproduce crucial features of gauge theories. 
We expect that the strategy allows the extension of iPEPS studies to more general LGTs.

\end{abstract}

\author{Daniel Robaina}
\affiliation{Max-Planck-Institut f\"ur Quantenoptik, Hans-Kopfermann-Str.\ 1, D-85748 Garching, Germany}
\author{Mari Carmen Ba\~nuls}
\affiliation{Max-Planck-Institut f\"ur Quantenoptik, Hans-Kopfermann-Str.\ 1, D-85748 Garching, Germany}
\affiliation{Munich Center for Quantum Science and Technology (MCQST), Schellingstr. 4, D-80799 M\"unchen}
\author{J. Ignacio Cirac}
\affiliation{Max-Planck-Institut f\"ur Quantenoptik, Hans-Kopfermann-Str.\ 1, D-85748 Garching, Germany}
\affiliation{Munich Center for Quantum Science and Technology (MCQST), Schellingstr. 4, D-80799 M\"unchen}
\date{\today}							
\maketitle

\paragraph{Introduction.--} 

For years, Tensor Networks (TN) have been exploited to study quantum many-body problems, especially in the context of condensed matter physics, since they provide efficient
 ans\"atze for ground states, low lying excitations and thermal equilibrium states of local
hamiltonians \cite{Cirac2009rg,Verstraete2008,Schollwoeck2011,Orus2013kga,Silvi2019tns}. The application of TN to Lattice Gauge Theories (LGT) constitutes a much newer, but also fast growing field. Their suitability for 1+1 dimensional problems has already been widely demonstrated using the matrix product state (MPS) ansatz.
In numerous studies, MPS have been shown to efficiently and accurately describe the relevant equilibrium physics of abelian and non-abelian LGTs, even at finite density where the infamous sign-problem would turn traditional Monte Carlo approaches infeasible, TN enable continuum limit extrapolations, as well as simulations in out-of-equilibrium scenarios (see \cite{Banuls:2019rao,qtflag2019} for recent reviews). 

The one-dimensional success strongly 
motivates an extension of the TN study to LGT in higher spatial dimensions, where the natural generalization of the MPS ansatz is provided by projected entangled pair states (PEPS) \cite{Verstraete:2004cf}, or its infinite version defined directly in the thermodynamic limit, iPEPS \cite{PhysRevLett.101.250602}. 
More restricted TN have allowed some first encouraging steps for two dimensional models. Early on, the phase diagram of a $\mathds{Z}_2$ LGT was studied with MERA\cite{Vidal2007,Tagliacozzo2011}, and, more recently,  tree tensor networks \cite{Shi2006} were applied to explore the $U(1)$ quantum link model on a finite lattice \cite{felser2019}.
But a fully variational PEPS calculation for a LGT does not yet exist.

Although the fast progress in iPEPS algorithms has allowed reaching some of the most competitive results for certain condensed matter problems \cite{Corboz2016,Corboz2016a,Vanderstraeten2016a,Corboz2018prx,Rader2018prx,Vanderstraeten2019exc,Hubig2019}
and there is no conceptual limitation to apply them to LGTs
\cite{Zapp2017}, until the date the only numerical results of (i)PEPS simulations of LGTs have been limited to toy models without an actual optimization of the most general tensors \cite{Tagliacozzo2014,Zohar2015c,Haegeman2015,Zohar:2016wcf,Zohar2018mc}.
Apart from the obvious increase in computational cost, another more limiting factor is the presence of plaquette terms in the LGT Hamiltonian. While it is possible to directly apply a plaquette term to PEPS \cite{Dusuel2011,Schulz2012topo}, this involves a considerably higher computational cost than the two-body interactions for which the most efficient PEPS algorithms are optimized, and ultimately limits the bond dimension that can be explored to only very small values, not enough to approach convergence.

In this work we develop a new update strategy that allows the standard plaquette term of a LGT to be applied as a sequence of purely two-body operations. This allows us to use an iPEPS ansatz to study the phase diagram of a $\mathds{Z}_3$-invariant LGT in two spatial dimensions.  In agreement to predictions in the literature \cite{KORTHALSALTES1978315, PhysRevLett.43.799,PhysRevD.21.2892}, we observe a confining and a non-confining phase.
We are able to quantitatively locate the transition at a value of the coupling constant $g^2_c = 1.159(4)$.
This constitutes the first ab initio iPEPS study of a 2+1d Lattice Gauge Theory,
and opens the door to studying a rich variety of LGTs using the most efficient up-to-date PEPS algorithms.

\paragraph{Model.--}
We consider a $\mathds{Z}_3$ invariant Lattice Gauge Theory given by the
following Hamiltonian in 2+1 space-time dimensions 

\begin{equation}
H = H_\textrm{E} + H_\square,
\label{eq:z3hamiltonian} 
\end{equation}
where
\begin{align}
H_\textrm{E} &= \frac{g^2}{2} \sum_{\bf x} E^2({\bf x}+{\bf i}/2) + E^2({\bf x}+{\bf j}/2)\nonumber \\
H_\square &= -\frac{1}{2g^2} \sum_{\bf x} U_P({\bf x}) + U^\dagger_P({\bf
x})\,.\nonumber
\end{align}
The plaquette operator is written as
\beq
U_P({\bf x}) = U^\dagger({\bf x }+{\bf j}/2) U^\dagger({\bf x} + {\bf i}/2 + {\bf j})
U({\bf x}+{\bf i} + {\bf j}/2) U({\bf x} + {\bf i}/2) \nonumber 
\eeq
where ${\bf x}$ is the position of a vertex and ${\bf i},{\bf j}$ are unit-vectors in both space directions  connecting two adjacent vertices. 

The physical degrees of freedom are the link variables which have a local
Hilbert space of dimension $d=3$ and consequently $E$ takes values in $\{-1,0,1\}$. The
unitary operators $U$ and $U^\dagger$, lower and raise respectively the electric
field at the corresponding link by one unit 
\begin{align}
U\ket{e} &= \ket{e-1}\nonumber\\
U^\dagger\ket{e} &= \ket{e +1}\nonumber
\end{align}
and $\mathds{Z}_3$-symmetry implies $U^3$ = $(U^\dagger)^3 = \mathds{1}$.

In the limit of $d \to \infty$ this $\mathds{Z}_d$ Hamiltonian yields a $U(1)$ Lattice Gauge Theory where $H_\textrm{E}$ corresponds to the
electric field and the plaquette terms in $H_\square$ reproduce the magnetic
parts \footnote{The limit of U(1) is recovered when $d\to \infty$ if the Hamiltonian is written in the form of \cite{Horn:1979fy} but for $d=3$ our formulation is equivalent except for a trivial rescaling of
$g^2$ and the $U$ operator and a constant overall shift in the Hamiltonian.}.

The Hamiltonian in \eqref{eq:z3hamiltonian} commutes with the Gauss-law operator
$G({\bf x})$ at every point in space giving rise to a local $\mathds{Z}_3$ gauge
symmetry where $G({\bf x})$ is given by

\beq
G({\bf x}) = e^{\frac{2\pi i}{3} (E_l({\bf x}) + E_d({\bf x}) -
E_r({\bf x}) - E_u({\bf x}))}
\eeq

where the subscripts $l,d,r,u$ correspond to the links which are to the \emph{left,
down, right, up} of the vertex at position ${\bf x}$. Notice that $G({\bf
x})$ is defined at the vertices of the lattice while the links live inbetween
vertices. Given that $[G({\bf x}),H] = 0$, the hamiltonian is block diagonal and
physical states that satisfy the Gauss-law obey 
\beq 
G({\bf x})\ket{\psi} = e^{\frac{2\pi i}{3} q({\bf x})} \ket{\psi},
\eeq
where $q({\bf x}) \in \{-1,0,1\}$ can be thought of as the static charge at
vertex ${\bf x}$. Although the ground state of the system lives in the charge
sector with $q({\bf x}) = 0$, $\forall {\bf x}$, it is also interesting to study
different charge patterns, as we will do.

\paragraph{Method.--}

An iPEPS ansatz consists of a unit-cell of rank-5 tensors arranged in a 2D-grid which is
repeated in both space directions infinitely many times. Those tensors have a physical index of
dimension $d$ equal to that of the local Hilbert space of each degree of freedom (3 in our case)
and 4 additional virtual indices of bond dimension $D$ that allow for the interactions
with neighbouring tensors. As $D$ increases the ansatz becomes more general and, consequently, a better description of the true quantum state is expected. 

There are several ways of optimizing the tensors within the unit-cell in order to find the ground state. One possibility relies on a
variational approach in which only one tensor is varied at a time by keeping the rest fixed.
The optimal tensor is then found by solving a Generalized Eigenvalue Problem before moving
to the next one \cite{Verstraete2008}. While the variational method has been able to obtain very accurate energies~\cite{Corboz2016a,Vanderstraeten2016a}, the most widely used strategy for iPEPS, which we also adopt here, is still an
 imaginary time evolution, very much in the spirit of the popular Time
Evolving Block Decimation (TEBD) algorithm \cite{PhysRevLett.91.147902}.
In the most efficient version, a
simple update (SU) \cite{PhysRevLett.101.090603} strategy is used to find the optimized tensors. 

We use a second order Suzuki-Trotter \cite{Trotter1959,Suzuki1985} expansion of the Hamiltonian exponential
\begin{equation}
e^{-\beta (H_E + H_\square)} = \lim_{n\to \infty} \left(e^{-\frac{\delta_\tau}{2} H_E}  e^{-\delta_\tau 
H_\square}e^{-\frac{\delta_\tau}{2} H_E}\right)^n  
\end{equation}

with $ \delta_\tau = \beta/n$ and $\beta$ the total imaginary time evolved until convergence.

Traditional iPEPS algorithms have been optimized for Hamiltonians with nearest neighbor interactions. Longer range or higher-order terms considerably increase the computational cost. Therefore, in order to apply these methods to our problem, we need a simple and efficient update strategy that takes into account 4-body plaquette operators like the ones that appear in LGTs.

\begin{figure}
\includegraphics[width=0.7\columnwidth]{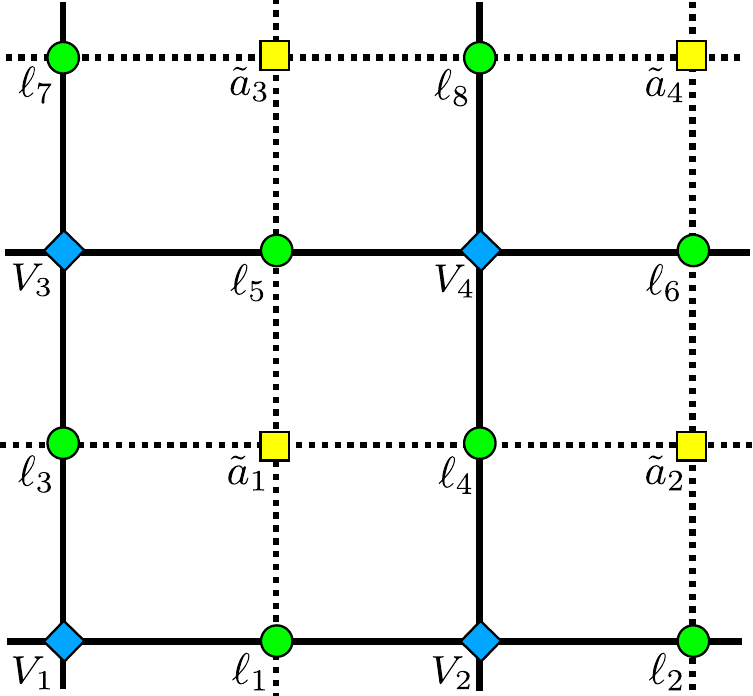}
\caption{iPEPS unit-cell.}
\label{fig:unit-cell}
\end{figure}

In order to apply the plaquette operator in its exponential form we import an idea originally envisioned for digital quantum simulations of LGTs \cite{Zohar:2016wmo,Zohar:2016iic,Zohar:2017hzz,Bender:2018rdp}. The key aspect
consists in including an auxiliary degree of freedom with the same Hilbert space
as the links themselves at the center of each plaquette. This ancilla is
prepared in a state which is an equal weight symmetric superposition of all basis states. Following the notation of \cite{Zohar:2016iic} we call it
$\ket{\widetilde{\rm in}} = \frac{1}{\sqrt{3}} \sum_{m=-1,0,1} \ket{\tilde{m}}$. The derivation
presented in the above mentioned papers allows us to write the action
of the four-body operator $e^{-\delta_\tau H_\square}$ as a sequence of two-body
gates (we call this the \emph{entangler}) followed by a local operation on the
ancilla. The inverse of the entangler (the \emph{disentangler}) 
leaves the ancilla back in its original state $\ket{\widetilde{\rm in}}$,  ready for the next update. The
full identity reads
\beq
\mathcal{U}^\dagger_\square e^{\frac{\delta \tau}{2g^2}(\tilde{U} + \tilde{U}^\dagger)} \mathcal{U}_\square
\ket{\widetilde{\rm in}} = \ket{\widetilde{\rm in}} e^{-\delta \tau H_\square} \,
\label{eq:stator_id}
\eeq
where the entangler $\mathcal{U}_\square = \mathcal{U}^\dagger_l
\mathcal{U}^\dagger_u \mathcal{U}_r \mathcal{U}_d$ is the product of four two-body gates
between ancilla and the corresponding links.
Each of these
two-body gates is written as
\beq 
    \mathcal{U}_i = U_i \otimes \widetilde{P}_1 + \mathds{1}_i \otimes \widetilde{P}_0 + U^\dagger_i 
\otimes \widetilde{P}_{-1}\,
\eeq
where $U_i$ with $i=l,u,r,d$ act on the links and $\widetilde{P}_m$ are ordinary projectors in the ancilla Hilbert space that
project onto state $\ket{\tilde{m}}$. The local operation on the ancilla
$e^{\frac{\delta \tau}{2g^2}(\widetilde{U} + \widetilde{U}^\dagger)}$
involves $\widetilde{U}$ and $\widetilde{U}^\dagger$ which are nothing but ordinary $U$ (and
$U^\dagger$)-operators acting on the ancilla degrees of freedom. Note, that \eqref{eq:stator_id} is a mathematical identity and there is no approximation involved. We refer the
interested reader to the original papers for a clean derivation of
\eqref{eq:stator_id}.

The electrical evolution corresponds to a sequential action of
$e^{-\frac{\delta_{\tau}g^2}{4} E^2}$-single-site operators onto the physical indices of all
links. Since we employ the simple update procedure (SU) this operation does not increase the bond dimensions and thus carries no truncation errors. 

In order to implement the  update procedure described above, we choose a $4\times 4$ unit cell as our iPEPS ansatz as shown in
Fig.~\ref{fig:unit-cell}.
The unit cell contains 16 different tensors, 8 of them corresponding to the gauge degrees of freedom residing on the links (green
circles labelled $\ell_i$, with $i=1,...,8$), plus four tensors for the ancillas (yellow squares) at the center of
the plaquettes and four for the vertices (blue diamonds). 
The solid lines represent the physical lattice of the system
that connects links and vertices while the dashed lines correspond to an
auxiliary lattice that connects ancillas with links.

iPEPS are able to account for global and local symmetries of the
theory by imposing a particular block structure of the tensors \cite{PerezGarcia2008sym,Singh2010,PerezGarcia2010njp,Tagliacozzo2014,Haegeman2015,Zohar2015b,Zohar2015c}.
In our case, this is ensured by applying a gauge projector that enforces the Gauss-law on the vertices \footnote{See the Supp.~Mat. for more details on the projection.}.
Since all the terms in the Hamiltonian commute with $G({\bf x})$, it is enough to apply the projector at the beginning of the imaginary time evolution. To cope with potential errors introduced by the truncation, we subsequently monitor the expectation value of $G({\bf x})$ to be sure to stay in the sector of interest. 
We observe that the deviation (with respect to the desired sector) is not larger than $10^{-6}$ in any of our simulations.

Similarly to
other Tensor Networks, iPEPS allow for the calculation of local observables. 
This requires an accurate approximation of
 the environment around a given tensor. In
this work 
we calculate the environment with the Corner Transfer Matrix (CTM)-method
\cite{Orus2009ctm,PhysRevB.82.245119},
which introduces an additional bond dimension, controlling the precision of such approximation \footnote{
An error analysis on the convergence of the CTM can be found in the Supp.~Mat.}.

Altogether,  this strategy allows us to simulate the imaginary time evolution of a LGT including the four-body plaquette operator by means of well-known tools to the iPEPS practitioners like single and two-body gates.
\paragraph{Phase Diagram.--}
When $g^2 \to \infty$, the electric field term dominates and, in the case of
vanishing static charges at all the vertices, the lowest energy is attained when all links are in the zero electric flux state. The ground state thus becomes a product state with zero energy. Similarly, in the weak coupling
regime when $g^2 \to 0$ the energy per plaquette tends towards the asymptotic
value of $-1/g^2$ where the ground state is again a product state. 
It is well known that $\mathds{Z}_d$ gauge theories are dual to spin systems with
nearest neighbour interactions \cite{KORTHALSALTES1978315}. For $\mathds{Z}_3$ in
$2+1$ dimensions the system undergoes a first order phase transition \cite{PhysRevLett.43.799, PhysRevD.21.2892} around some critical coupling
$g^2_c$. 

\begin{figure}
    \includegraphics[width=\columnwidth]{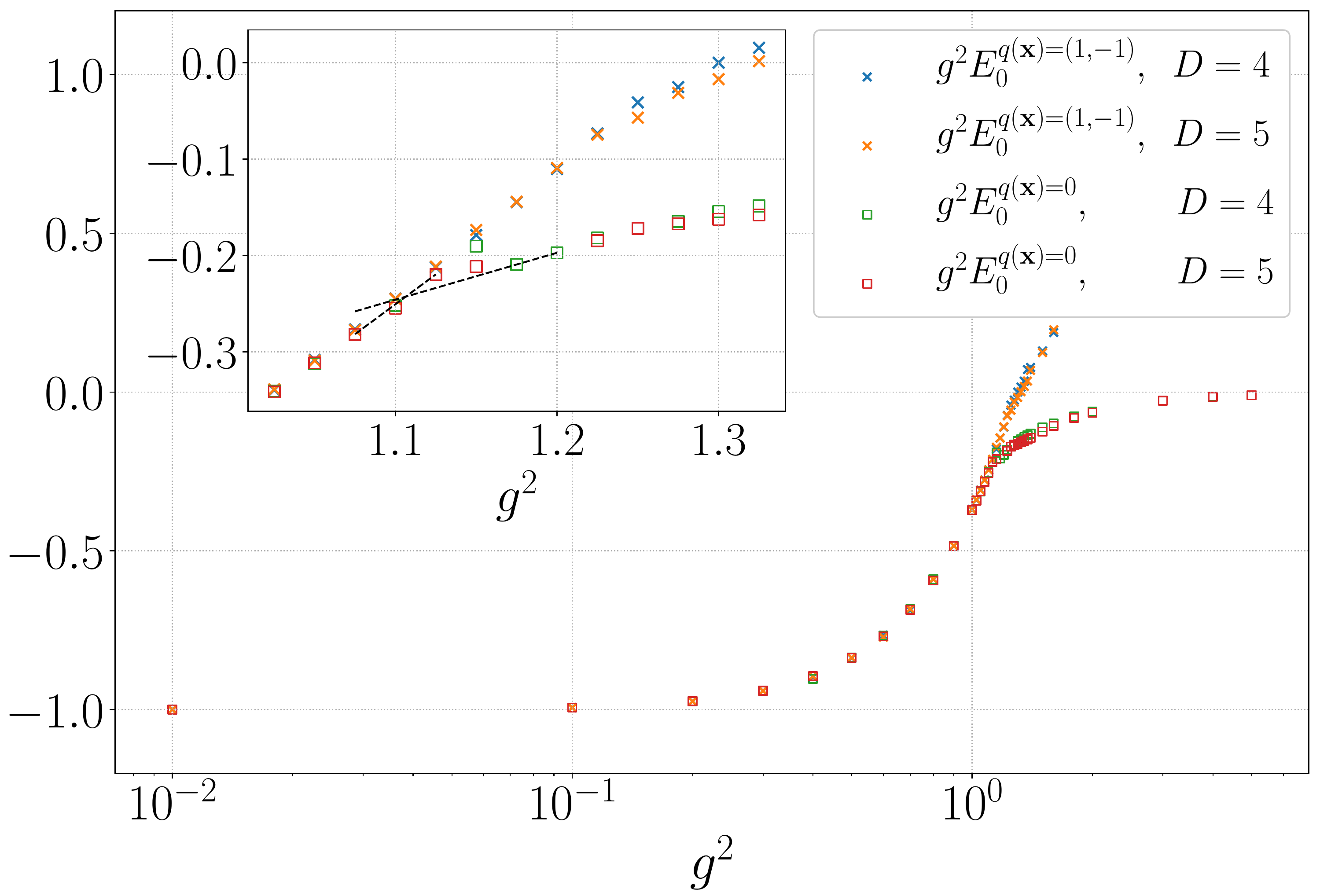}
    \caption{Ground State Energies for the zero charge sector with bond dimensions $D=4$, $5$.
    We compare to the sector of two adjacent vertices respectively projected to charges 1 and -1 with bond dimension $D=4,5$. Inset: Transition region
    zoom in. }
    \label{fig:energyfig}
\end{figure}

We have performed calculations at $D=3$, $4$, $5$ for the whole range of couplings
from $g^2 = 0.01$ to $g^2 = 5.0$.  
As expected, increasing the bond dimension yields lower energies in general. We observe that for some values of the coupling constant near the phase transition,
$D=5$ was not able to provide a lower estimate than $D=4$.
We attribute this to a lack of full convergence of the SU on those points. Since for the rest of parameters the relative difference between the results for $D=4$ and $5$ is extremely  small (see SM),
we take $D=4$ as our best data-set and use $D=3$ and $5$ to estimate
numerical errors \footnote{See Supp. Mat. for an error estimate on our data.}. Our ground-state energy results are plotted in \figref{fig:energyfig}. 

\begin{figure}
    \includegraphics[width=\columnwidth]{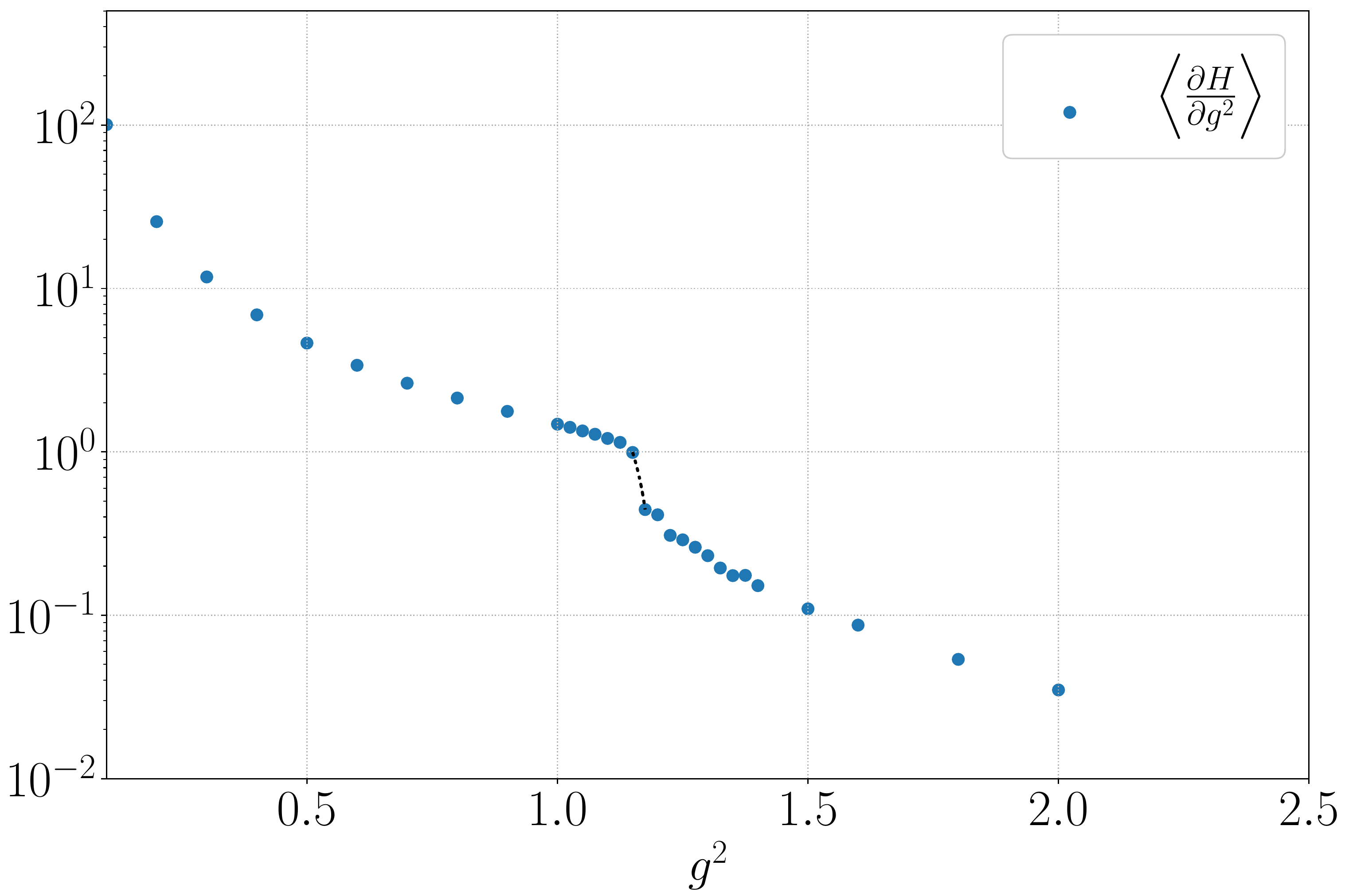}
    \caption{Expectation value of $\frac{\partial H}{\partial g^2}$ on the ground state for
    the zero charge sector. Bond dimension is $D=4$.}
    \label{fig:derivfig}
\end{figure}

First order phase transitions can be cleanly detected by TN simulations \cite{PhysRevLett.102.077203}
as cusps in the energy curve, corresponding to a level crossing. This effect is apparent in \figref{fig:energyfig} at intermediate values of the coupling (the dashed lines are meant to guide the eye). A cleaner way
of locating the phase transition is by
the discontinuity in the first derivative of the energy, which can be calculated as
\beq
\frac{\partial E_0^{q({\bf x})=0}}{\partial g^2} =
\left<\psi_\textrm{GS}\right|\frac{\partial{H}}{\partial g^2}\left|\psi_\textrm{GS}\right>
= \frac{1}{g^2} \left<\psi_\textrm{GS}\right|H_\textrm{E}-H_\square\left|\psi_\textrm{GS}\right>
\eeq
and is plotted in \figref{fig:derivfig}. A clear discontinuity between $g^2_c=1.15$ and
$g^2_c=1.175$ can be identified. 

\begin{figure}
    \includegraphics[width=\columnwidth]{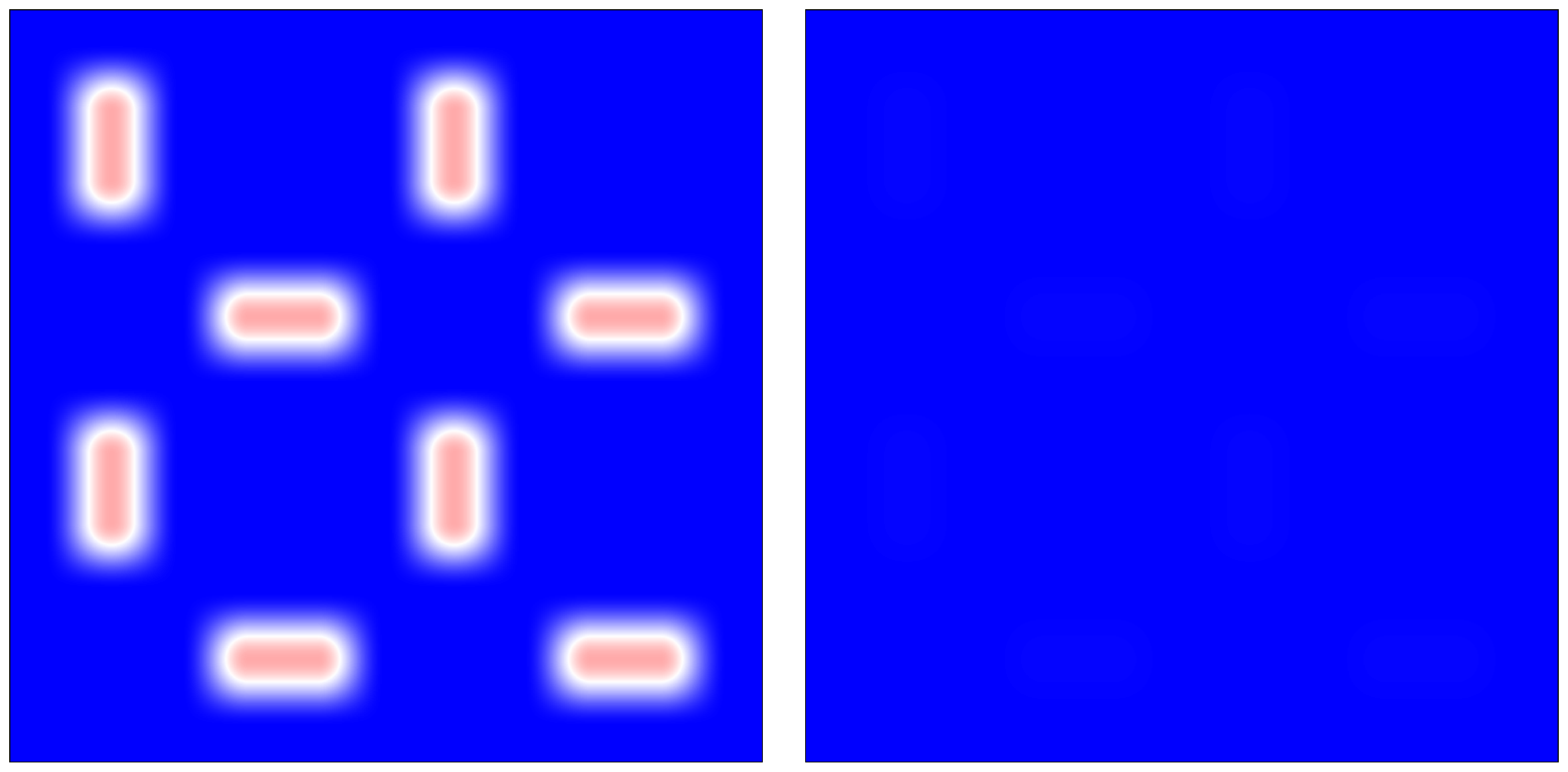}
    \includegraphics[width=\columnwidth]{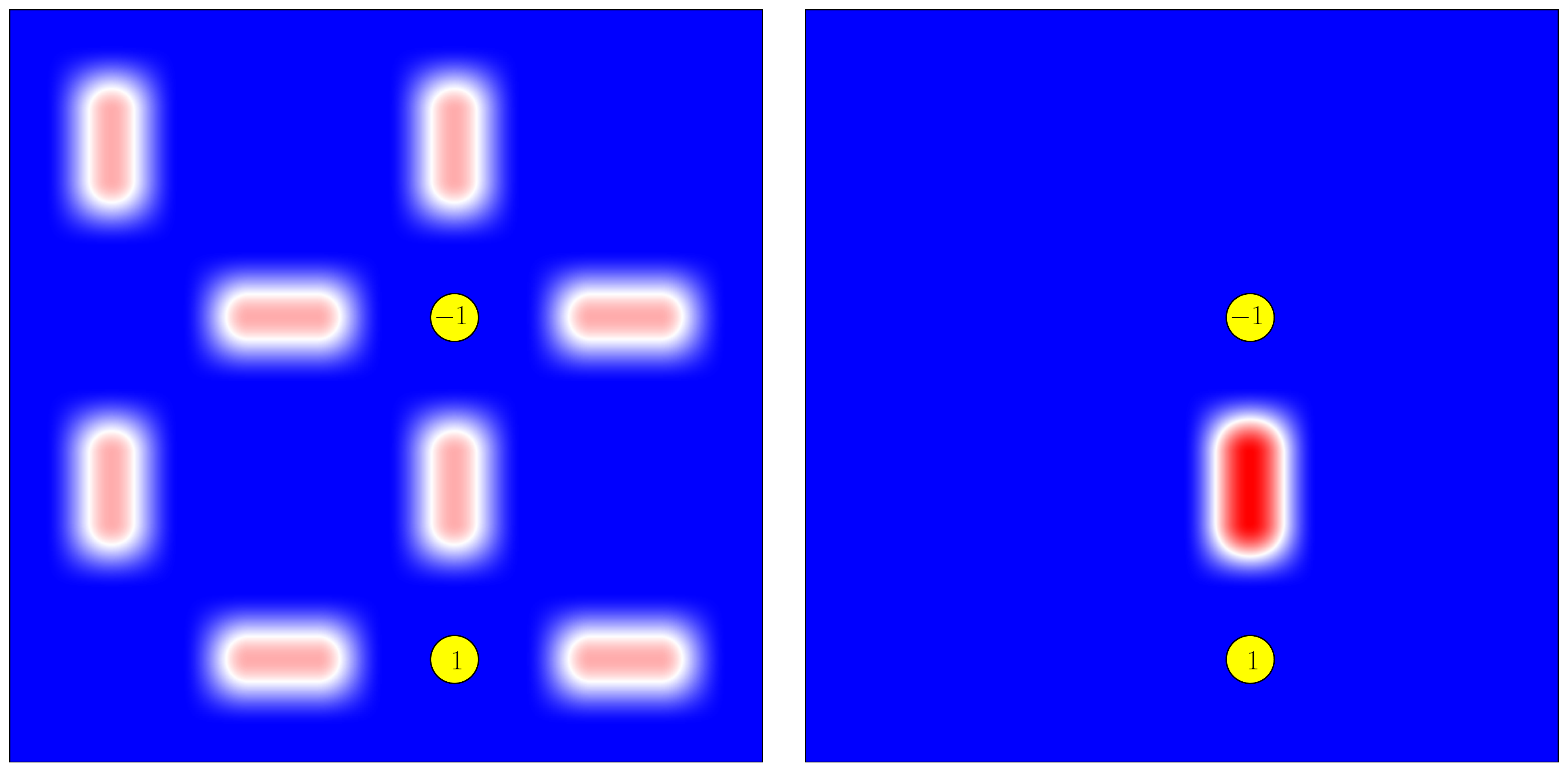}
    \caption{Ground state expectation values of $E^2$-operators acting on the links of the
    unit-cell for $g^2=0.01,5.0$. The upper row corresponds
    to the zero static charge sector while the lower has two vertices (yellow
    circles) projected to $1$ and $-1$ static charges.}
    \label{fig:Emaps}
\end{figure}

We also consider a different charge sector, in which we project
two adjacent vertices to static charges 1 and -1 respectively (as illustrated in
\figref{fig:Emaps}). 
Below the phase transition, both sectors are close to degenerate (see
\figref{fig:energyfig}),
and as soon the transition is crossed, they separate.  The energy per plaquette of the
static charges tends to $g^2/8$ in the limit of $g^2 \to \infty$ since our unit cell
contains 4 plaquettes and in that limit there is a single link whose $E^2$-expectation
value is 1, while the rest vanish. The fact that the energies of both sectors start to
strongly deviate from each other exactly at the phase transition represents a consistency
check that we have correctly located the transition region. We will attempt a more
accurate determination of $g^2_c$ via Wilson loops in the following section.

\begin{figure}
\includegraphics[width=\columnwidth]{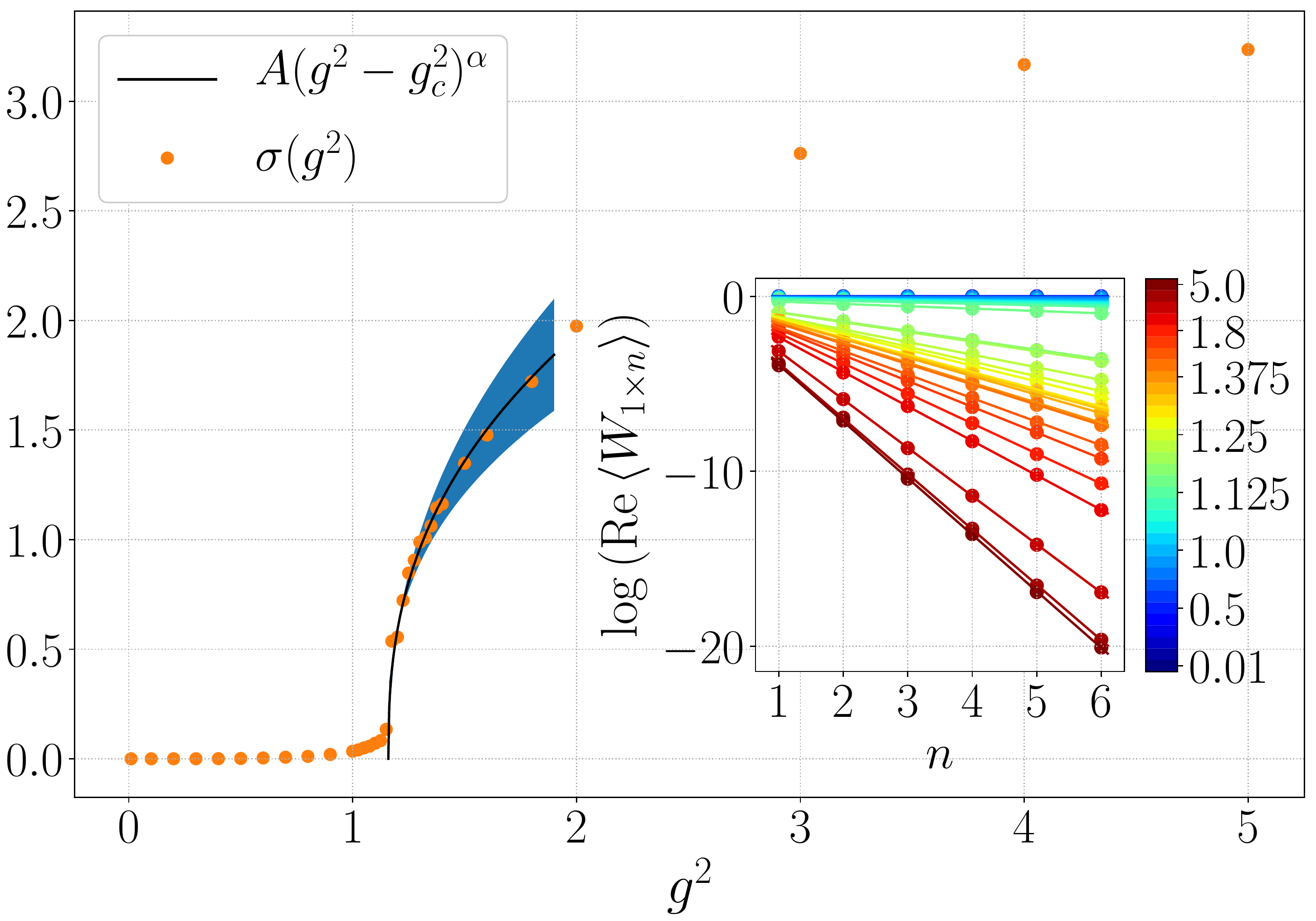}
\caption{Area-law coefficient $\sigma$ obtained from the fit of the expectation value of the Wilson
loops that is shown in the inset for $D=4$ ground states. The colorbar represents
the value of the coupling $g^2$. The blue band represents an error estimation for the fitted curve.}
\label{fig:area-law}
\end{figure}

\paragraph{Wilson loops.--}
The phase transition separates a non-confining (for small $g^2$) from a confining (for large $g^2$) phase. We can characterize it by investigating the ground state expectation value of several closed
spatial Wilson loops, the simplest of them being the plaquette which enters the
calculation of the energy. In the confining phase, these values are expected to decay exponentially with the area of the loop.
Due to the large computational cost of these
quantities, we restrict ourselves to loops of width 1 and length $n=1,\,\ldots 6$.
The corresponding operator can be written in closed form as

\begin{align}
& W_{1\times n} = U^\dagger ({\bf x} + {\bf j}/2) \otimes
\left(\bigotimes^{n-1}_{\alpha=0}
U({\bf x} + (\alpha + 1/2) {\bf i})\right) \nonumber\\
&\otimes U({\bf x} + n{\bf i} + {\bf j}/2) \otimes
\left( \bigotimes^{n-1}_{\beta=0}U^\dagger({\bf x} + (n-\beta - 1/2){\bf i} +
{\bf j})\right)\,.
\end{align}
We calculate $\left<\psi_\textrm{GS}\right|W_{1\times
n}\left|\psi_\textrm{GS}\right>$ and show the results in figure
\ref{fig:area-law}. 
We perform a linear fit of the logarithm of the real part of
$\left<W_{1\times n}\right>$ (the imaginary part is consistent with zero) vs. the area $n$, and read off the slope $\sigma$. 
The phase transition clearly manifests in a sudden increase of $\sigma$ when the coupling
approaches a critical value $g^2_c$. In order to extract this critical value, we perform several fits of the data to a form
$A(g^2-g^2_c)^\alpha$ and estimate the errors by varying the number of points included in
the fit. We find 
\begin{equation}
A=2.0(3), \qquad g^2_c = 1.159(4), \qquad \alpha = 0.39(3)\,.
\end{equation}

\paragraph{Electric field map.--}
In order to illustrate clearly the very different behavior of the electric field in both
phases, in figure \ref{fig:Emaps} we plot $\left<\psi_\textrm{GS}\right|E^2_\ell\left|\psi_\textrm{GS}\right>$
for all 8 links in the unit-cell in different charge sectors. The zero charge sector keeps translational symmetry for
all values of the coupling and above the phase transition the electric field is
practically zero. For the case of two static charges, we see that below the phase
transition the behavior is very similar as in the zero charge sector, while as soon as the
transition is crossed, the electric field is confined to a single link between two charges. 

\paragraph{Conclusions.--}

We find that iPEPS are capable of accurately capturing the Physics of a gauge theory with different phases in 2+1 space-time dimensions.
With moderate bond dimension, the iPEPS ansatz allows us not only to determine the ground state energy but also to explore the phenomenology of the model, including the location of a confinement phase transition.

Key to this development is a special update strategy that employs additional ancillary degrees of freedom and reduces many-body terms to sequences of two-body operations.
This allows us to deal with plaquette terms in an efficient way, and also to correctly implement Gauss-law constraints at the vertices as a way to impose the local symmetry.

The strategy can be immediately applied to other LGTs, but also to other hamiltonians that require the inclusion of a 4-body operator. 
Since the original construction \cite{Bender:2018rdp} on which this update is based can be applied to non-Abelian Lie groups and also to operators acting on a larger number of sites~\cite{Zohar:2017hzz,Zohar2020local}, we expect that the method can be further generalized.
Dynamical fermions can additionally be included in the approach without involving a sign-problem, and we leave this direction for future work.
Altogether, this opens the door to more ambitious iPEPS studies of LGTs.

\acknowledgments{

We thank Claudius Hubig for insightful discussions on the SyTen toolkit used in this work
\cite{hubig:_syten_toolk,hubig17:_symmet_protec_tensor_networ}.
This work was partly supported by the Deutsche Forschungsgemeinschaft (DFG, German
Research Foundation)under Germany's Excellence Strategy -- EXC-2111 -- 390814868,  and EU-QUANTERA project QTFLAG (BMBF grant No. 13N14780).
}

\bibliographystyle{apsrev4-1}
\bibliography{Z3biblio}

\begin{thebibliography}{53}%
\makeatletter
\providecommand \@ifxundefined [1]{%
 \@ifx{#1\undefined}
}%
\providecommand \@ifnum [1]{%
 \ifnum #1\expandafter \@firstoftwo
 \else \expandafter \@secondoftwo
 \fi
}%
\providecommand \@ifx [1]{%
 \ifx #1\expandafter \@firstoftwo
 \else \expandafter \@secondoftwo
 \fi
}%
\providecommand \natexlab [1]{#1}%
\providecommand \enquote  [1]{``#1''}%
\providecommand \bibnamefont  [1]{#1}%
\providecommand \bibfnamefont [1]{#1}%
\providecommand \citenamefont [1]{#1}%
\providecommand \href@noop [0]{\@secondoftwo}%
\providecommand \href [0]{\begingroup \@sanitize@url \@href}%
\providecommand \@href[1]{\@@startlink{#1}\@@href}%
\providecommand \@@href[1]{\endgroup#1\@@endlink}%
\providecommand \@sanitize@url [0]{\catcode `\\12\catcode `\$12\catcode
  `\&12\catcode `\#12\catcode `\^12\catcode `\_12\catcode `\%12\relax}%
\providecommand \@@startlink[1]{}%
\providecommand \@@endlink[0]{}%
\providecommand \url  [0]{\begingroup\@sanitize@url \@url }%
\providecommand \@url [1]{\endgroup\@href {#1}{\urlprefix }}%
\providecommand \urlprefix  [0]{URL }%
\providecommand \Eprint [0]{\href }%
\providecommand \doibase [0]{http://dx.doi.org/}%
\providecommand \selectlanguage [0]{\@gobble}%
\providecommand \bibinfo  [0]{\@secondoftwo}%
\providecommand \bibfield  [0]{\@secondoftwo}%
\providecommand \translation [1]{[#1]}%
\providecommand \BibitemOpen [0]{}%
\providecommand \bibitemStop [0]{}%
\providecommand \bibitemNoStop [0]{.\EOS\space}%
\providecommand \EOS [0]{\spacefactor3000\relax}%
\providecommand \BibitemShut  [1]{\csname bibitem#1\endcsname}%
\let\auto@bib@innerbib\@empty
\bibitem [{\citenamefont {Cirac}\ and\ \citenamefont
  {Verstraete}(2009)}]{Cirac2009rg}%
  \BibitemOpen
  \bibfield  {author} {\bibinfo {author} {\bibfnamefont {J.~I.}\ \bibnamefont
  {Cirac}}\ and\ \bibinfo {author} {\bibfnamefont {F.}~\bibnamefont
  {Verstraete}},\ }\href {http://stacks.iop.org/1751-8121/42/i=50/a=504004}
  {\bibfield  {journal} {\bibinfo  {journal} {Journal of Physics A:
  Mathematical and Theoretical}\ }\textbf {\bibinfo {volume} {42}},\ \bibinfo
  {pages} {504004} (\bibinfo {year} {2009})}\BibitemShut {NoStop}%
\bibitem [{\citenamefont {Verstraete}\ \emph {et~al.}(2008)\citenamefont
  {Verstraete}, \citenamefont {Murg},\ and\ \citenamefont
  {Cirac}}]{Verstraete2008}%
  \BibitemOpen
  \bibfield  {author} {\bibinfo {author} {\bibfnamefont {F.}~\bibnamefont
  {Verstraete}}, \bibinfo {author} {\bibfnamefont {V.}~\bibnamefont {Murg}}, \
  and\ \bibinfo {author} {\bibfnamefont {J.}~\bibnamefont {Cirac}},\ }\href
  {\doibase 10.1080/14789940801912366} {\bibfield  {journal} {\bibinfo
  {journal} {Adv. Phys.}\ }\textbf {\bibinfo {volume} {57}},\ \bibinfo {pages}
  {143} (\bibinfo {year} {2008})}\BibitemShut {NoStop}%
\bibitem [{\citenamefont {Schollw\"{o}ck}(2011)}]{Schollwoeck2011}%
  \BibitemOpen
  \bibfield  {author} {\bibinfo {author} {\bibfnamefont {U.}~\bibnamefont
  {Schollw\"{o}ck}},\ }\href {\doibase 10.1016/j.aop.2010.09.012} {\bibfield
  {journal} {\bibinfo  {journal} {Ann. Phys.}\ }\textbf {\bibinfo {volume}
  {326}},\ \bibinfo {pages} {96 } (\bibinfo {year} {2011})},\ \bibinfo {note}
  {january 2011 Special Issue}\BibitemShut {NoStop}%
\bibitem [{\citenamefont {Or{\'u}s}(2014)}]{Orus2013kga}%
  \BibitemOpen
  \bibfield  {author} {\bibinfo {author} {\bibfnamefont {R.}~\bibnamefont
  {Or{\'u}s}},\ }\href {\doibase 10.1016/j.aop.2014.06.013} {\bibfield
  {journal} {\bibinfo  {journal} {Annals Phys.}\ }\textbf {\bibinfo {volume}
  {349}},\ \bibinfo {pages} {117} (\bibinfo {year} {2014})},\ \Eprint
  {http://arxiv.org/abs/1306.2164} {arXiv:1306.2164 [cond-mat.str-el]}
  \BibitemShut {NoStop}%
\bibitem [{\citenamefont {Silvi}\ \emph {et~al.}(2019)\citenamefont {Silvi},
  \citenamefont {Tschirsich}, \citenamefont {Gerster}, \citenamefont
  {J{\"u}nemann}, \citenamefont {Jaschke}, \citenamefont {Rizzi},\ and\
  \citenamefont {Montangero}}]{Silvi2019tns}%
  \BibitemOpen
  \bibfield  {author} {\bibinfo {author} {\bibfnamefont {P.}~\bibnamefont
  {Silvi}}, \bibinfo {author} {\bibfnamefont {F.}~\bibnamefont {Tschirsich}},
  \bibinfo {author} {\bibfnamefont {M.}~\bibnamefont {Gerster}}, \bibinfo
  {author} {\bibfnamefont {J.}~\bibnamefont {J{\"u}nemann}}, \bibinfo {author}
  {\bibfnamefont {D.}~\bibnamefont {Jaschke}}, \bibinfo {author} {\bibfnamefont
  {M.}~\bibnamefont {Rizzi}}, \ and\ \bibinfo {author} {\bibfnamefont
  {S.}~\bibnamefont {Montangero}},\ }\href {\doibase
  10.21468/SciPostPhysLectNotes.8} {\bibfield  {journal} {\bibinfo  {journal}
  {SciPost Phys. Lect. Notes}\ ,\ \bibinfo {pages} {8}} (\bibinfo {year}
  {2019})}\BibitemShut {NoStop}%
\bibitem [{\citenamefont {Ba{\~n}uls}\ and\ \citenamefont
  {Cichy}(2020)}]{Banuls:2019rao}%
  \BibitemOpen
  \bibfield  {author} {\bibinfo {author} {\bibfnamefont {M.~C.}\ \bibnamefont
  {Ba{\~n}uls}}\ and\ \bibinfo {author} {\bibfnamefont {K.}~\bibnamefont
  {Cichy}},\ }\href {\doibase 10.1088/1361-6633/ab6311} {\bibfield  {journal}
  {\bibinfo  {journal} {Rept. Prog. Phys.}\ }\textbf {\bibinfo {volume} {83}},\
  \bibinfo {pages} {024401} (\bibinfo {year} {2020})},\ \Eprint
  {http://arxiv.org/abs/1910.00257} {arXiv:1910.00257 [hep-lat]} \BibitemShut
  {NoStop}%
\bibitem [{\citenamefont {{Ba{\~n}uls}}\ \emph {et~al.}(2019)\citenamefont
  {{Ba{\~n}uls}}, \citenamefont {{Blatt}}, \citenamefont {{Catani}},
  \citenamefont {{Celi}}, \citenamefont {{Cirac}}, \citenamefont {{Dalmonte}},
  \citenamefont {{Fallani}}, \citenamefont {{Jansen}}, \citenamefont
  {{Lewenstein}}, \citenamefont {{Montangero}}, \citenamefont {{Muschik}},
  \citenamefont {{Reznik}}, \citenamefont {{Rico}}, \citenamefont
  {{Tagliacozzo}}, \citenamefont {{Van Acoleyen}}, \citenamefont
  {{Verstraete}}, \citenamefont {{Wiese}}, \citenamefont {{Wingate}},
  \citenamefont {{Zakrzewski}},\ and\ \citenamefont {{Zoller}}}]{qtflag2019}%
  \BibitemOpen
  \bibfield  {author} {\bibinfo {author} {\bibfnamefont {M.~C.}\ \bibnamefont
  {{Ba{\~n}uls}}}, \bibinfo {author} {\bibfnamefont {R.}~\bibnamefont
  {{Blatt}}}, \bibinfo {author} {\bibfnamefont {J.}~\bibnamefont {{Catani}}},
  \bibinfo {author} {\bibfnamefont {A.}~\bibnamefont {{Celi}}}, \bibinfo
  {author} {\bibfnamefont {J.~I.}\ \bibnamefont {{Cirac}}}, \bibinfo {author}
  {\bibfnamefont {M.}~\bibnamefont {{Dalmonte}}}, \bibinfo {author}
  {\bibfnamefont {L.}~\bibnamefont {{Fallani}}}, \bibinfo {author}
  {\bibfnamefont {K.}~\bibnamefont {{Jansen}}}, \bibinfo {author}
  {\bibfnamefont {M.}~\bibnamefont {{Lewenstein}}}, \bibinfo {author}
  {\bibfnamefont {S.}~\bibnamefont {{Montangero}}}, \bibinfo {author}
  {\bibfnamefont {C.~A.}\ \bibnamefont {{Muschik}}}, \bibinfo {author}
  {\bibfnamefont {B.}~\bibnamefont {{Reznik}}}, \bibinfo {author}
  {\bibfnamefont {E.}~\bibnamefont {{Rico}}}, \bibinfo {author} {\bibfnamefont
  {L.}~\bibnamefont {{Tagliacozzo}}}, \bibinfo {author} {\bibfnamefont
  {K.}~\bibnamefont {{Van Acoleyen}}}, \bibinfo {author} {\bibfnamefont
  {F.}~\bibnamefont {{Verstraete}}}, \bibinfo {author} {\bibfnamefont {U.~J.}\
  \bibnamefont {{Wiese}}}, \bibinfo {author} {\bibfnamefont {M.}~\bibnamefont
  {{Wingate}}}, \bibinfo {author} {\bibfnamefont {J.}~\bibnamefont
  {{Zakrzewski}}}, \ and\ \bibinfo {author} {\bibfnamefont {P.}~\bibnamefont
  {{Zoller}}},\ }\href@noop {} {\bibfield  {journal} {\bibinfo  {journal}
  {arXiv e-prints}\ ,\ \bibinfo {eid} {arXiv:1911.00003}} (\bibinfo {year}
  {2019})},\ \Eprint {http://arxiv.org/abs/1911.00003} {arXiv:1911.00003
  [quant-ph]} \BibitemShut {NoStop}%
\bibitem [{\citenamefont {Verstraete}\ and\ \citenamefont
  {Cirac}(2004)}]{Verstraete:2004cf}%
  \BibitemOpen
  \bibfield  {author} {\bibinfo {author} {\bibfnamefont {F.}~\bibnamefont
  {Verstraete}}\ and\ \bibinfo {author} {\bibfnamefont {J.}~\bibnamefont
  {Cirac}},\ }\href@noop {} {\  (\bibinfo {year} {2004})},\ \Eprint
  {http://arxiv.org/abs/cond-mat/0407066} {arXiv:cond-mat/0407066} \BibitemShut
  {NoStop}%
\bibitem [{\citenamefont {Jordan}\ \emph {et~al.}(2008)\citenamefont {Jordan},
  \citenamefont {Or\'us}, \citenamefont {Vidal}, \citenamefont {Verstraete},\
  and\ \citenamefont {Cirac}}]{PhysRevLett.101.250602}%
  \BibitemOpen
  \bibfield  {author} {\bibinfo {author} {\bibfnamefont {J.}~\bibnamefont
  {Jordan}}, \bibinfo {author} {\bibfnamefont {R.}~\bibnamefont {Or\'us}},
  \bibinfo {author} {\bibfnamefont {G.}~\bibnamefont {Vidal}}, \bibinfo
  {author} {\bibfnamefont {F.}~\bibnamefont {Verstraete}}, \ and\ \bibinfo
  {author} {\bibfnamefont {J.~I.}\ \bibnamefont {Cirac}},\ }\href {\doibase
  10.1103/PhysRevLett.101.250602} {\bibfield  {journal} {\bibinfo  {journal}
  {Phys. Rev. Lett.}\ }\textbf {\bibinfo {volume} {101}},\ \bibinfo {pages}
  {250602} (\bibinfo {year} {2008})}\BibitemShut {NoStop}%
\bibitem [{\citenamefont {Vidal}(2007)}]{Vidal2007}%
  \BibitemOpen
  \bibfield  {author} {\bibinfo {author} {\bibfnamefont {G.}~\bibnamefont
  {Vidal}},\ }\href {\doibase 10.1103/PhysRevLett.99.220405} {\bibfield
  {journal} {\bibinfo  {journal} {Phys. Rev. Lett.}\ }\textbf {\bibinfo
  {volume} {99}},\ \bibinfo {pages} {220405} (\bibinfo {year}
  {2007})}\BibitemShut {NoStop}%
\bibitem [{\citenamefont {Tagliacozzo}\ and\ \citenamefont
  {Vidal}(2011)}]{Tagliacozzo2011}%
  \BibitemOpen
  \bibfield  {author} {\bibinfo {author} {\bibfnamefont {L.}~\bibnamefont
  {Tagliacozzo}}\ and\ \bibinfo {author} {\bibfnamefont {G.}~\bibnamefont
  {Vidal}},\ }\href {\doibase 10.1103/PhysRevB.83.115127} {\bibfield  {journal}
  {\bibinfo  {journal} {Phys. Rev. B}\ }\textbf {\bibinfo {volume} {83}},\
  \bibinfo {pages} {115127} (\bibinfo {year} {2011})},\ \Eprint
  {http://arxiv.org/abs/arxiv:1007.4145} {arxiv:1007.4145} \BibitemShut
  {NoStop}%
\bibitem [{\citenamefont {Shi}\ \emph {et~al.}(2006)\citenamefont {Shi},
  \citenamefont {Duan},\ and\ \citenamefont {Vidal}}]{Shi2006}%
  \BibitemOpen
  \bibfield  {author} {\bibinfo {author} {\bibfnamefont {Y.-Y.}\ \bibnamefont
  {Shi}}, \bibinfo {author} {\bibfnamefont {L.-M.}\ \bibnamefont {Duan}}, \
  and\ \bibinfo {author} {\bibfnamefont {G.}~\bibnamefont {Vidal}},\ }\href
  {\doibase 10.1103/PhysRevA.74.022320} {\bibfield  {journal} {\bibinfo
  {journal} {Phys. Rev. A}\ }\textbf {\bibinfo {volume} {74}},\ \bibinfo
  {pages} {022320} (\bibinfo {year} {2006})}\BibitemShut {NoStop}%
\bibitem [{\citenamefont {Felser}\ \emph {et~al.}(2019)\citenamefont {Felser},
  \citenamefont {Silvi}, \citenamefont {Collura},\ and\ \citenamefont
  {Montangero}}]{felser2019}%
  \BibitemOpen
  \bibfield  {author} {\bibinfo {author} {\bibfnamefont {T.}~\bibnamefont
  {Felser}}, \bibinfo {author} {\bibfnamefont {P.}~\bibnamefont {Silvi}},
  \bibinfo {author} {\bibfnamefont {M.}~\bibnamefont {Collura}}, \ and\
  \bibinfo {author} {\bibfnamefont {S.}~\bibnamefont {Montangero}},\
  }\href@noop {} {\enquote {\bibinfo {title} {Two-dimensional quantum-link
  lattice quantum electrodynamics at finite density},}\ } (\bibinfo {year}
  {2019})\BibitemShut {NoStop}%
\bibitem [{\citenamefont {Corboz}(2016{\natexlab{a}})}]{Corboz2016}%
  \BibitemOpen
  \bibfield  {author} {\bibinfo {author} {\bibfnamefont {P.}~\bibnamefont
  {Corboz}},\ }\href {\doibase 10.1103/PhysRevB.93.045116} {\bibfield
  {journal} {\bibinfo  {journal} {Phys. Rev. B}\ }\textbf {\bibinfo {volume}
  {93}},\ \bibinfo {pages} {045116} (\bibinfo {year}
  {2016}{\natexlab{a}})}\BibitemShut {NoStop}%
\bibitem [{\citenamefont {Corboz}(2016{\natexlab{b}})}]{Corboz2016a}%
  \BibitemOpen
  \bibfield  {author} {\bibinfo {author} {\bibfnamefont {P.}~\bibnamefont
  {Corboz}},\ }\href {\doibase 10.1103/PhysRevB.94.035133} {\bibfield
  {journal} {\bibinfo  {journal} {Phys. Rev. B}\ }\textbf {\bibinfo {volume}
  {94}},\ \bibinfo {pages} {035133} (\bibinfo {year}
  {2016}{\natexlab{b}})}\BibitemShut {NoStop}%
\bibitem [{\citenamefont {Vanderstraeten}\ \emph {et~al.}(2016)\citenamefont
  {Vanderstraeten}, \citenamefont {Haegeman}, \citenamefont {Corboz},\ and\
  \citenamefont {Verstraete}}]{Vanderstraeten2016a}%
  \BibitemOpen
  \bibfield  {author} {\bibinfo {author} {\bibfnamefont {L.}~\bibnamefont
  {Vanderstraeten}}, \bibinfo {author} {\bibfnamefont {J.}~\bibnamefont
  {Haegeman}}, \bibinfo {author} {\bibfnamefont {P.}~\bibnamefont {Corboz}}, \
  and\ \bibinfo {author} {\bibfnamefont {F.}~\bibnamefont {Verstraete}},\
  }\href {\doibase 10.1103/PhysRevB.94.155123} {\bibfield  {journal} {\bibinfo
  {journal} {Phys. Rev. B}\ }\textbf {\bibinfo {volume} {94}},\ \bibinfo
  {pages} {155123} (\bibinfo {year} {2016})}\BibitemShut {NoStop}%
\bibitem [{\citenamefont {Corboz}\ \emph {et~al.}(2018)\citenamefont {Corboz},
  \citenamefont {Czarnik}, \citenamefont {Kapteijns},\ and\ \citenamefont
  {Tagliacozzo}}]{Corboz2018prx}%
  \BibitemOpen
  \bibfield  {author} {\bibinfo {author} {\bibfnamefont {P.}~\bibnamefont
  {Corboz}}, \bibinfo {author} {\bibfnamefont {P.}~\bibnamefont {Czarnik}},
  \bibinfo {author} {\bibfnamefont {G.}~\bibnamefont {Kapteijns}}, \ and\
  \bibinfo {author} {\bibfnamefont {L.}~\bibnamefont {Tagliacozzo}},\ }\href
  {\doibase 10.1103/PhysRevX.8.031031} {\bibfield  {journal} {\bibinfo
  {journal} {Phys. Rev. X}\ }\textbf {\bibinfo {volume} {8}},\ \bibinfo {pages}
  {031031} (\bibinfo {year} {2018})}\BibitemShut {NoStop}%
\bibitem [{\citenamefont {Rader}\ and\ \citenamefont
  {L{\"a}uchli}(2018)}]{Rader2018prx}%
  \BibitemOpen
  \bibfield  {author} {\bibinfo {author} {\bibfnamefont {M.}~\bibnamefont
  {Rader}}\ and\ \bibinfo {author} {\bibfnamefont {A.~M.}\ \bibnamefont
  {L{\"a}uchli}},\ }\href {\doibase 10.1103/PhysRevX.8.031030} {\bibfield
  {journal} {\bibinfo  {journal} {Phys. Rev. X}\ }\textbf {\bibinfo {volume}
  {8}},\ \bibinfo {pages} {031030} (\bibinfo {year} {2018})}\BibitemShut
  {NoStop}%
\bibitem [{\citenamefont {Vanderstraeten}\ \emph {et~al.}(2019)\citenamefont
  {Vanderstraeten}, \citenamefont {Haegeman},\ and\ \citenamefont
  {Verstraete}}]{Vanderstraeten2019exc}%
  \BibitemOpen
  \bibfield  {author} {\bibinfo {author} {\bibfnamefont {L.}~\bibnamefont
  {Vanderstraeten}}, \bibinfo {author} {\bibfnamefont {J.}~\bibnamefont
  {Haegeman}}, \ and\ \bibinfo {author} {\bibfnamefont {F.}~\bibnamefont
  {Verstraete}},\ }\href {\doibase 10.1103/PhysRevB.99.165121} {\bibfield
  {journal} {\bibinfo  {journal} {Phys. Rev. B}\ }\textbf {\bibinfo {volume}
  {99}},\ \bibinfo {pages} {165121} (\bibinfo {year} {2019})}\BibitemShut
  {NoStop}%
\bibitem [{\citenamefont {Hubig}\ and\ \citenamefont
  {Cirac}(2019)}]{Hubig2019}%
  \BibitemOpen
  \bibfield  {author} {\bibinfo {author} {\bibfnamefont {C.}~\bibnamefont
  {Hubig}}\ and\ \bibinfo {author} {\bibfnamefont {J.~I.}\ \bibnamefont
  {Cirac}},\ }\href {\doibase 10.21468/SciPostPhys.6.3.031} {\bibfield
  {journal} {\bibinfo  {journal} {SciPost Phys.}\ }\textbf {\bibinfo {volume}
  {6}},\ \bibinfo {pages} {31} (\bibinfo {year} {2019})}\BibitemShut {NoStop}%
\bibitem [{\citenamefont {Zapp}\ and\ \citenamefont {Or\'us}(2017)}]{Zapp2017}%
  \BibitemOpen
  \bibfield  {author} {\bibinfo {author} {\bibfnamefont {K.}~\bibnamefont
  {Zapp}}\ and\ \bibinfo {author} {\bibfnamefont {R.}~\bibnamefont {Or\'us}},\
  }\href {\doibase 10.1103/PhysRevD.95.114508} {\bibfield  {journal} {\bibinfo
  {journal} {Phys. Rev. D}\ }\textbf {\bibinfo {volume} {95}},\ \bibinfo
  {pages} {114508} (\bibinfo {year} {2017})}\BibitemShut {NoStop}%
\bibitem [{\citenamefont {Tagliacozzo}\ \emph {et~al.}(2014)\citenamefont
  {Tagliacozzo}, \citenamefont {Celi},\ and\ \citenamefont
  {Lewenstein}}]{Tagliacozzo2014}%
  \BibitemOpen
  \bibfield  {author} {\bibinfo {author} {\bibfnamefont {L.}~\bibnamefont
  {Tagliacozzo}}, \bibinfo {author} {\bibfnamefont {A.}~\bibnamefont {Celi}}, \
  and\ \bibinfo {author} {\bibfnamefont {M.}~\bibnamefont {Lewenstein}},\
  }\href {\doibase 10.1103/PhysRevX.4.041024} {\bibfield  {journal} {\bibinfo
  {journal} {Phys. Rev. X}\ }\textbf {\bibinfo {volume} {4}},\ \bibinfo {pages}
  {041024} (\bibinfo {year} {2014})}\BibitemShut {NoStop}%
\bibitem [{\citenamefont {Zohar}\ \emph {et~al.}(2015)\citenamefont {Zohar},
  \citenamefont {Burrello}, \citenamefont {Wahl},\ and\ \citenamefont
  {Cirac}}]{Zohar2015c}%
  \BibitemOpen
  \bibfield  {author} {\bibinfo {author} {\bibfnamefont {E.}~\bibnamefont
  {Zohar}}, \bibinfo {author} {\bibfnamefont {M.}~\bibnamefont {Burrello}},
  \bibinfo {author} {\bibfnamefont {T.~B.}\ \bibnamefont {Wahl}}, \ and\
  \bibinfo {author} {\bibfnamefont {J.~I.}\ \bibnamefont {Cirac}},\ }\href
  {\doibase 10.1016/j.aop.2015.10.009} {\bibfield  {journal} {\bibinfo
  {journal} {Ann. Phys. (Amsterdam)}\ }\textbf {\bibinfo {volume} {363}},\
  \bibinfo {pages} {385 } (\bibinfo {year} {2015})}\BibitemShut {NoStop}%
\bibitem [{\citenamefont {Haegeman}\ \emph {et~al.}(2015)\citenamefont
  {Haegeman}, \citenamefont {Van~Acoleyen}, \citenamefont {Schuch},
  \citenamefont {Cirac},\ and\ \citenamefont {Verstraete}}]{Haegeman2015}%
  \BibitemOpen
  \bibfield  {author} {\bibinfo {author} {\bibfnamefont {J.}~\bibnamefont
  {Haegeman}}, \bibinfo {author} {\bibfnamefont {K.}~\bibnamefont
  {Van~Acoleyen}}, \bibinfo {author} {\bibfnamefont {N.}~\bibnamefont
  {Schuch}}, \bibinfo {author} {\bibfnamefont {J.~I.}\ \bibnamefont {Cirac}}, \
  and\ \bibinfo {author} {\bibfnamefont {F.}~\bibnamefont {Verstraete}},\
  }\href {\doibase 10.1103/PhysRevX.5.011024} {\bibfield  {journal} {\bibinfo
  {journal} {Phys. Rev. X}\ }\textbf {\bibinfo {volume} {5}},\ \bibinfo {pages}
  {011024} (\bibinfo {year} {2015})}\BibitemShut {NoStop}%
\bibitem [{\citenamefont {Zohar}\ \emph {et~al.}(2016)\citenamefont {Zohar},
  \citenamefont {Wahl}, \citenamefont {Burrello},\ and\ \citenamefont
  {Cirac}}]{Zohar:2016wcf}%
  \BibitemOpen
  \bibfield  {author} {\bibinfo {author} {\bibfnamefont {E.}~\bibnamefont
  {Zohar}}, \bibinfo {author} {\bibfnamefont {T.~B.}\ \bibnamefont {Wahl}},
  \bibinfo {author} {\bibfnamefont {M.}~\bibnamefont {Burrello}}, \ and\
  \bibinfo {author} {\bibfnamefont {J.~I.}\ \bibnamefont {Cirac}},\ }\href
  {\doibase 10.1016/j.aop.2016.08.008} {\bibfield  {journal} {\bibinfo
  {journal} {Annals Phys.}\ }\textbf {\bibinfo {volume} {374}},\ \bibinfo
  {pages} {84} (\bibinfo {year} {2016})},\ \Eprint
  {http://arxiv.org/abs/1607.08115} {arXiv:1607.08115 [quant-ph]} \BibitemShut
  {NoStop}%
\bibitem [{\citenamefont {Zohar}\ and\ \citenamefont
  {Cirac}(2018)}]{Zohar2018mc}%
  \BibitemOpen
  \bibfield  {author} {\bibinfo {author} {\bibfnamefont {E.}~\bibnamefont
  {Zohar}}\ and\ \bibinfo {author} {\bibfnamefont {J.~I.}\ \bibnamefont
  {Cirac}},\ }\href {\doibase 10.1103/PhysRevD.97.034510} {\bibfield  {journal}
  {\bibinfo  {journal} {Phys. Rev. D}\ }\textbf {\bibinfo {volume} {97}},\
  \bibinfo {pages} {034510} (\bibinfo {year} {2018})}\BibitemShut {NoStop}%
\bibitem [{\citenamefont {Dusuel}\ \emph {et~al.}(2011)\citenamefont {Dusuel},
  \citenamefont {Kamfor}, \citenamefont {Or\'us}, \citenamefont {Schmidt},\
  and\ \citenamefont {Vidal}}]{Dusuel2011}%
  \BibitemOpen
  \bibfield  {author} {\bibinfo {author} {\bibfnamefont {S.}~\bibnamefont
  {Dusuel}}, \bibinfo {author} {\bibfnamefont {M.}~\bibnamefont {Kamfor}},
  \bibinfo {author} {\bibfnamefont {R.}~\bibnamefont {Or\'us}}, \bibinfo
  {author} {\bibfnamefont {K.~P.}\ \bibnamefont {Schmidt}}, \ and\ \bibinfo
  {author} {\bibfnamefont {J.}~\bibnamefont {Vidal}},\ }\href {\doibase
  10.1103/PhysRevLett.106.107203} {\bibfield  {journal} {\bibinfo  {journal}
  {Phys. Rev. Lett.}\ }\textbf {\bibinfo {volume} {106}},\ \bibinfo {pages}
  {107203} (\bibinfo {year} {2011})}\BibitemShut {NoStop}%
\bibitem [{\citenamefont {Schulz}\ \emph {et~al.}(2012)\citenamefont {Schulz},
  \citenamefont {Dusuel}, \citenamefont {Or{\'{u}}s}, \citenamefont {Vidal},\
  and\ \citenamefont {Schmidt}}]{Schulz2012topo}%
  \BibitemOpen
  \bibfield  {author} {\bibinfo {author} {\bibfnamefont {M.~D.}\ \bibnamefont
  {Schulz}}, \bibinfo {author} {\bibfnamefont {S.}~\bibnamefont {Dusuel}},
  \bibinfo {author} {\bibfnamefont {R.}~\bibnamefont {Or{\'{u}}s}}, \bibinfo
  {author} {\bibfnamefont {J.}~\bibnamefont {Vidal}}, \ and\ \bibinfo {author}
  {\bibfnamefont {K.~P.}\ \bibnamefont {Schmidt}},\ }\href {\doibase
  10.1088/1367-2630/14/2/025005} {\bibfield  {journal} {\bibinfo  {journal}
  {New Journal of Physics}\ }\textbf {\bibinfo {volume} {14}},\ \bibinfo
  {pages} {025005} (\bibinfo {year} {2012})}\BibitemShut {NoStop}%
\bibitem [{\citenamefont {Altes}(1978)}]{KORTHALSALTES1978315}%
  \BibitemOpen
  \bibfield  {author} {\bibinfo {author} {\bibfnamefont {C.~P.~K.}\
  \bibnamefont {Altes}},\ }\href {\doibase
  https://doi.org/10.1016/0550-3213(78)90207-9} {\bibfield  {journal} {\bibinfo
   {journal} {Nuclear Physics B}\ }\textbf {\bibinfo {volume} {142}},\ \bibinfo
  {pages} {315 } (\bibinfo {year} {1978})}\BibitemShut {NoStop}%
\bibitem [{\citenamefont {Bl\"ote}\ and\ \citenamefont
  {Swendsen}(1979)}]{PhysRevLett.43.799}%
  \BibitemOpen
  \bibfield  {author} {\bibinfo {author} {\bibfnamefont {H.~W.~J.}\
  \bibnamefont {Bl\"ote}}\ and\ \bibinfo {author} {\bibfnamefont {R.~H.}\
  \bibnamefont {Swendsen}},\ }\href {\doibase 10.1103/PhysRevLett.43.799}
  {\bibfield  {journal} {\bibinfo  {journal} {Phys. Rev. Lett.}\ }\textbf
  {\bibinfo {volume} {43}},\ \bibinfo {pages} {799} (\bibinfo {year}
  {1979})}\BibitemShut {NoStop}%
\bibitem [{\citenamefont {Bhanot}\ and\ \citenamefont
  {Creutz}(1980)}]{PhysRevD.21.2892}%
  \BibitemOpen
  \bibfield  {author} {\bibinfo {author} {\bibfnamefont {G.}~\bibnamefont
  {Bhanot}}\ and\ \bibinfo {author} {\bibfnamefont {M.}~\bibnamefont
  {Creutz}},\ }\href {\doibase 10.1103/PhysRevD.21.2892} {\bibfield  {journal}
  {\bibinfo  {journal} {Phys. Rev. D}\ }\textbf {\bibinfo {volume} {21}},\
  \bibinfo {pages} {2892} (\bibinfo {year} {1980})}\BibitemShut {NoStop}%
\bibitem [{Note1()}]{Note1}%
  \BibitemOpen
  \bibinfo {note} {The limit of U(1) is recovered when $d\to \infty $ if the
  Hamiltonian is written in the form of \cite {Horn:1979fy} but for $d=3$ our
  formulation is equivalent except for a trivial rescaling of $g^2$ and the $U$
  operator and a constant overall shift in the Hamiltonian.}\BibitemShut
  {Stop}%
\bibitem [{\citenamefont {Vidal}(2003)}]{PhysRevLett.91.147902}%
  \BibitemOpen
  \bibfield  {author} {\bibinfo {author} {\bibfnamefont {G.}~\bibnamefont
  {Vidal}},\ }\href {\doibase 10.1103/PhysRevLett.91.147902} {\bibfield
  {journal} {\bibinfo  {journal} {Phys. Rev. Lett.}\ }\textbf {\bibinfo
  {volume} {91}},\ \bibinfo {pages} {147902} (\bibinfo {year}
  {2003})}\BibitemShut {NoStop}%
\bibitem [{\citenamefont {Jiang}\ \emph {et~al.}(2008)\citenamefont {Jiang},
  \citenamefont {Weng},\ and\ \citenamefont {Xiang}}]{PhysRevLett.101.090603}%
  \BibitemOpen
  \bibfield  {author} {\bibinfo {author} {\bibfnamefont {H.~C.}\ \bibnamefont
  {Jiang}}, \bibinfo {author} {\bibfnamefont {Z.~Y.}\ \bibnamefont {Weng}}, \
  and\ \bibinfo {author} {\bibfnamefont {T.}~\bibnamefont {Xiang}},\ }\href
  {\doibase 10.1103/PhysRevLett.101.090603} {\bibfield  {journal} {\bibinfo
  {journal} {Phys. Rev. Lett.}\ }\textbf {\bibinfo {volume} {101}},\ \bibinfo
  {pages} {090603} (\bibinfo {year} {2008})}\BibitemShut {NoStop}%
\bibitem [{\citenamefont {Trotter}(1959)}]{Trotter1959}%
  \BibitemOpen
  \bibfield  {author} {\bibinfo {author} {\bibfnamefont {H.~F.}\ \bibnamefont
  {Trotter}},\ }\href@noop {} {\bibfield  {journal} {\bibinfo  {journal} {Proc.
  Amer. Math. Soc.}\ }\textbf {\bibinfo {volume} {10}},\ \bibinfo {pages} {545}
  (\bibinfo {year} {1959})}\BibitemShut {NoStop}%
\bibitem [{\citenamefont {Suzuki}(1985)}]{Suzuki1985}%
  \BibitemOpen
  \bibfield  {author} {\bibinfo {author} {\bibfnamefont {M.}~\bibnamefont
  {Suzuki}},\ }\href {\doibase 10.1063/1.526596} {\bibfield  {journal}
  {\bibinfo  {journal} {Journal of Mathematical Physics}\ }\textbf {\bibinfo
  {volume} {26}},\ \bibinfo {pages} {601} (\bibinfo {year} {1985})}\BibitemShut
  {NoStop}%
\bibitem [{\citenamefont {Zohar}\ \emph
  {et~al.}(2017{\natexlab{a}})\citenamefont {Zohar}, \citenamefont {Farace},
  \citenamefont {Reznik},\ and\ \citenamefont {Cirac}}]{Zohar:2016wmo}%
  \BibitemOpen
  \bibfield  {author} {\bibinfo {author} {\bibfnamefont {E.}~\bibnamefont
  {Zohar}}, \bibinfo {author} {\bibfnamefont {A.}~\bibnamefont {Farace}},
  \bibinfo {author} {\bibfnamefont {B.}~\bibnamefont {Reznik}}, \ and\ \bibinfo
  {author} {\bibfnamefont {J.~I.}\ \bibnamefont {Cirac}},\ }\href {\doibase
  10.1103/PhysRevLett.118.070501} {\bibfield  {journal} {\bibinfo  {journal}
  {Phys. Rev. Lett.}\ }\textbf {\bibinfo {volume} {118}},\ \bibinfo {pages}
  {070501} (\bibinfo {year} {2017}{\natexlab{a}})},\ \Eprint
  {http://arxiv.org/abs/1607.03656} {arXiv:1607.03656 [quant-ph]} \BibitemShut
  {NoStop}%
\bibitem [{\citenamefont {Zohar}\ \emph
  {et~al.}(2017{\natexlab{b}})\citenamefont {Zohar}, \citenamefont {Farace},
  \citenamefont {Reznik},\ and\ \citenamefont {Cirac}}]{Zohar:2016iic}%
  \BibitemOpen
  \bibfield  {author} {\bibinfo {author} {\bibfnamefont {E.}~\bibnamefont
  {Zohar}}, \bibinfo {author} {\bibfnamefont {A.}~\bibnamefont {Farace}},
  \bibinfo {author} {\bibfnamefont {B.}~\bibnamefont {Reznik}}, \ and\ \bibinfo
  {author} {\bibfnamefont {J.~I.}\ \bibnamefont {Cirac}},\ }\href {\doibase
  10.1103/PhysRevA.95.023604} {\bibfield  {journal} {\bibinfo  {journal} {Phys.
  Rev.}\ }\textbf {\bibinfo {volume} {A95}},\ \bibinfo {pages} {023604}
  (\bibinfo {year} {2017}{\natexlab{b}})},\ \Eprint
  {http://arxiv.org/abs/1607.08121} {arXiv:1607.08121 [quant-ph]} \BibitemShut
  {NoStop}%
\bibitem [{\citenamefont {Zohar}(2017)}]{Zohar:2017hzz}%
  \BibitemOpen
  \bibfield  {author} {\bibinfo {author} {\bibfnamefont {E.}~\bibnamefont
  {Zohar}},\ }\href {\doibase 10.1088/1751-8121/aa55ef} {\bibfield  {journal}
  {\bibinfo  {journal} {J. Phys.}\ }\textbf {\bibinfo {volume} {A50}},\
  \bibinfo {pages} {085301} (\bibinfo {year} {2017})},\ \Eprint
  {http://arxiv.org/abs/1607.08122} {arXiv:1607.08122 [quant-ph]} \BibitemShut
  {NoStop}%
\bibitem [{\citenamefont {Bender}\ \emph {et~al.}(2018)\citenamefont {Bender},
  \citenamefont {Zohar}, \citenamefont {Farace},\ and\ \citenamefont
  {Cirac}}]{Bender:2018rdp}%
  \BibitemOpen
  \bibfield  {author} {\bibinfo {author} {\bibfnamefont {J.}~\bibnamefont
  {Bender}}, \bibinfo {author} {\bibfnamefont {E.}~\bibnamefont {Zohar}},
  \bibinfo {author} {\bibfnamefont {A.}~\bibnamefont {Farace}}, \ and\ \bibinfo
  {author} {\bibfnamefont {J.~I.}\ \bibnamefont {Cirac}},\ }\href {\doibase
  10.1088/1367-2630/aadb71} {\bibfield  {journal} {\bibinfo  {journal} {New J.
  Phys.}\ }\textbf {\bibinfo {volume} {20}},\ \bibinfo {pages} {093001}
  (\bibinfo {year} {2018})},\ \Eprint {http://arxiv.org/abs/1804.02082}
  {arXiv:1804.02082 [quant-ph]} \BibitemShut {NoStop}%
\bibitem [{\citenamefont {P{\'e}rez-Garc\'{\i}a}\ \emph
  {et~al.}(2008)\citenamefont {P{\'e}rez-Garc\'{\i}a}, \citenamefont {Wolf},
  \citenamefont {Sanz}, \citenamefont {Verstraete},\ and\ \citenamefont
  {Cirac}}]{PerezGarcia2008sym}%
  \BibitemOpen
  \bibfield  {author} {\bibinfo {author} {\bibfnamefont {D.}~\bibnamefont
  {P{\'e}rez-Garc\'{\i}a}}, \bibinfo {author} {\bibfnamefont {M.~M.}\
  \bibnamefont {Wolf}}, \bibinfo {author} {\bibfnamefont {M.}~\bibnamefont
  {Sanz}}, \bibinfo {author} {\bibfnamefont {F.}~\bibnamefont {Verstraete}}, \
  and\ \bibinfo {author} {\bibfnamefont {J.~I.}\ \bibnamefont {Cirac}},\ }\href
  {\doibase 10.1103/PhysRevLett.100.167202} {\bibfield  {journal} {\bibinfo
  {journal} {Phys. Rev. Lett.}\ }\textbf {\bibinfo {volume} {100}},\ \bibinfo
  {pages} {167202} (\bibinfo {year} {2008})}\BibitemShut {NoStop}%
\bibitem [{\citenamefont {Singh}\ \emph {et~al.}(2010)\citenamefont {Singh},
  \citenamefont {Pfeifer},\ and\ \citenamefont {Vidal}}]{Singh2010}%
  \BibitemOpen
  \bibfield  {author} {\bibinfo {author} {\bibfnamefont {S.}~\bibnamefont
  {Singh}}, \bibinfo {author} {\bibfnamefont {R.~N.~C.}\ \bibnamefont
  {Pfeifer}}, \ and\ \bibinfo {author} {\bibfnamefont {G.}~\bibnamefont
  {Vidal}},\ }\href {\doibase 10.1103/PhysRevA.82.050301} {\bibfield  {journal}
  {\bibinfo  {journal} {Phys. Rev. A}\ }\textbf {\bibinfo {volume} {82}},\
  \bibinfo {pages} {050301} (\bibinfo {year} {2010})}\BibitemShut {NoStop}%
\bibitem [{\citenamefont {P{\'e}rez-Garc{\'\i}a}\ \emph
  {et~al.}(2010)\citenamefont {P{\'e}rez-Garc{\'\i}a}, \citenamefont {Sanz},
  \citenamefont {Gonz{\'a}lez-Guill{\'e}n}, \citenamefont {Wolf},\ and\
  \citenamefont {Cirac}}]{PerezGarcia2010njp}%
  \BibitemOpen
  \bibfield  {author} {\bibinfo {author} {\bibfnamefont {D.}~\bibnamefont
  {P{\'e}rez-Garc{\'\i}a}}, \bibinfo {author} {\bibfnamefont {M.}~\bibnamefont
  {Sanz}}, \bibinfo {author} {\bibfnamefont {C.~E.}\ \bibnamefont
  {Gonz{\'a}lez-Guill{\'e}n}}, \bibinfo {author} {\bibfnamefont {M.~M.}\
  \bibnamefont {Wolf}}, \ and\ \bibinfo {author} {\bibfnamefont {J.~I.}\
  \bibnamefont {Cirac}},\ }\href
  {http://stacks.iop.org/1367-2630/12/i=2/a=025010} {\bibfield  {journal}
  {\bibinfo  {journal} {New Journal of Physics}\ }\textbf {\bibinfo {volume}
  {12}},\ \bibinfo {pages} {025010} (\bibinfo {year} {2010})}\BibitemShut
  {NoStop}%
\bibitem [{\citenamefont {Zohar}\ and\ \citenamefont
  {Burrello}(2016)}]{Zohar2015b}%
  \BibitemOpen
  \bibfield  {author} {\bibinfo {author} {\bibfnamefont {E.}~\bibnamefont
  {Zohar}}\ and\ \bibinfo {author} {\bibfnamefont {M.}~\bibnamefont
  {Burrello}},\ }\href {\doibase 10.1088/1367-2630/18/4/043008} {\bibfield
  {journal} {\bibinfo  {journal} {New J. Phys.}\ }\textbf {\bibinfo {volume}
  {18}},\ \bibinfo {pages} {043008} (\bibinfo {year} {2016})}\BibitemShut
  {NoStop}%
\bibitem [{Note2()}]{Note2}%
  \BibitemOpen
  \bibinfo {note} {See the Supp.~Mat. for more details on the
  projection.}\BibitemShut {Stop}%
\bibitem [{\citenamefont {Or{\'u}s}\ and\ \citenamefont
  {Vidal}(2009)}]{Orus2009ctm}%
  \BibitemOpen
  \bibfield  {author} {\bibinfo {author} {\bibfnamefont {R.}~\bibnamefont
  {Or{\'u}s}}\ and\ \bibinfo {author} {\bibfnamefont {G.}~\bibnamefont
  {Vidal}},\ }\href {\doibase 10.1103/PhysRevB.80.094403} {\bibfield  {journal}
  {\bibinfo  {journal} {Phys. Rev. B}\ }\textbf {\bibinfo {volume} {80}},\
  \bibinfo {pages} {094403} (\bibinfo {year} {2009})}\BibitemShut {NoStop}%
\bibitem [{\citenamefont {Corboz}\ \emph {et~al.}(2010)\citenamefont {Corboz},
  \citenamefont {Jordan},\ and\ \citenamefont {Vidal}}]{PhysRevB.82.245119}%
  \BibitemOpen
  \bibfield  {author} {\bibinfo {author} {\bibfnamefont {P.}~\bibnamefont
  {Corboz}}, \bibinfo {author} {\bibfnamefont {J.}~\bibnamefont {Jordan}}, \
  and\ \bibinfo {author} {\bibfnamefont {G.}~\bibnamefont {Vidal}},\ }\href
  {\doibase 10.1103/PhysRevB.82.245119} {\bibfield  {journal} {\bibinfo
  {journal} {Phys. Rev. B}\ }\textbf {\bibinfo {volume} {82}},\ \bibinfo
  {pages} {245119} (\bibinfo {year} {2010})}\BibitemShut {NoStop}%
\bibitem [{Note3()}]{Note3}%
  \BibitemOpen
  \bibinfo {note} {An error analysis on the convergence of the CTM can be found
  in the Supp.~Mat.}\BibitemShut {Stop}%
\bibitem [{Note4()}]{Note4}%
  \BibitemOpen
  \bibinfo {note} {See Supp. Mat. for an error estimate on our
  data.}\BibitemShut {Stop}%
\bibitem [{\citenamefont {Or\'us}\ \emph {et~al.}(2009)\citenamefont {Or\'us},
  \citenamefont {Doherty},\ and\ \citenamefont
  {Vidal}}]{PhysRevLett.102.077203}%
  \BibitemOpen
  \bibfield  {author} {\bibinfo {author} {\bibfnamefont {R.}~\bibnamefont
  {Or\'us}}, \bibinfo {author} {\bibfnamefont {A.~C.}\ \bibnamefont {Doherty}},
  \ and\ \bibinfo {author} {\bibfnamefont {G.}~\bibnamefont {Vidal}},\ }\href
  {\doibase 10.1103/PhysRevLett.102.077203} {\bibfield  {journal} {\bibinfo
  {journal} {Phys. Rev. Lett.}\ }\textbf {\bibinfo {volume} {102}},\ \bibinfo
  {pages} {077203} (\bibinfo {year} {2009})}\BibitemShut {NoStop}%
\bibitem [{\citenamefont {Zohar}(2020)}]{Zohar2020local}%
  \BibitemOpen
  \bibfield  {author} {\bibinfo {author} {\bibfnamefont {E.}~\bibnamefont
  {Zohar}},\ }\href {\doibase 10.1103/PhysRevD.101.034518} {\bibfield
  {journal} {\bibinfo  {journal} {Phys. Rev. D}\ }\textbf {\bibinfo {volume}
  {101}},\ \bibinfo {pages} {034518} (\bibinfo {year} {2020})}\BibitemShut
  {NoStop}%
\bibitem [{hub()}]{hubig:_syten_toolk}%
  \BibitemOpen
  \href {https://syten.eu} {\enquote {\bibinfo {title} {The \textsc{SyTen}
  toolkit},}\ }\BibitemShut {NoStop}%
\bibitem [{\citenamefont {Hubig}(2017)}]{hubig17:_symmet_protec_tensor_networ}%
  \BibitemOpen
  \bibfield  {author} {\bibinfo {author} {\bibfnamefont {C.}~\bibnamefont
  {Hubig}},\ }\emph {\bibinfo {title} {Symmetry-Protected Tensor Networks}},\
  \href {https://edoc.ub.uni-muenchen.de/21348/} {Ph.D. thesis},\ \bibinfo
  {school} {LMU M{\"u}nchen} (\bibinfo {year} {2017})\BibitemShut {NoStop}%
\end{thebibliography}%
\twocolumngrid
\onecolumngrid

\section{Supplementary Material}
\subsection{Gauss-Law Constrains}
In order to enforce the Gauss-Law at every vertex, we define the projector 
\beq
P_{q}({\bf x}) = \frac{1}{3} \sum_{n=-1,0,1} \left(e^{\frac{2\pi i}{3}(E_l({\bf x}) + E_d({\bf x}) -
E_r({\bf x}) - E_u({\bf x})-q({\bf x})}\right)^n
\eeq
which projects vertex ${\bf x}$ to charge $q({\bf x})$. Since $E$-field operators in
the exponent commute with eachother, this projector has the same structure as
$H_\square$ since it can be written as a product of four single-site operators. Taking $q({\bf x})=0$ as an example case, it is convenient to consider a slight modification of identity \eqref{eq:stator_id}
\beq
\mathcal{G}^\dagger \frac{1}{3} \left(\tilde{\mathds{1}} + \tilde{U} +
\tilde{U}^\dagger\right) \mathcal{G}
\ket{\widetilde{\rm in}} = \ket{\widetilde{\rm in}} P_0 \,
\label{eq:proj_id}
\eeq
where now the entangler between vertex ${\bf x}$ and the links surrounding it can be again
written as a sequence of four two-body gates
$\mathcal{G} = \mathcal{G}_l \mathcal{G}_d \mathcal{G}^\dagger_r \mathcal{G}^\dagger_u$. 
Each of the two-body gates is written as
\beq 
    \mathcal{G}_i = g_i \otimes \tilde{P}_1 + \mathds{1}_i \otimes \tilde{P}_0 + g^\dagger_i 
\otimes \tilde{P}_{-1}\,
\eeq
with $g_i = e^{\frac{2\pi i }{3}E_i({\bf x})}$ and $i = l,u,r,d$. Similarly to the case with the ancillas, \eqref{eq:proj_id} is only true if vertex tensors are initialized in their
\ket{\widetilde{\rm in}} states. In this way, enforcing the Gauss-law at every vertex is as simple as applying a sequence of single and two-body gates. Only a minor modification to the local operation $\frac{1}{3} \left(\tilde{\mathds{1}} + \tilde{U} +
\tilde{U}^\dagger\right)$ on the vertex allows us to also obtain $P_q({\bf x})$ with $q({\bf x}) = \pm 1$.

\subsection{Errors}

\begin{figure}
\includegraphics[width=.49\columnwidth]{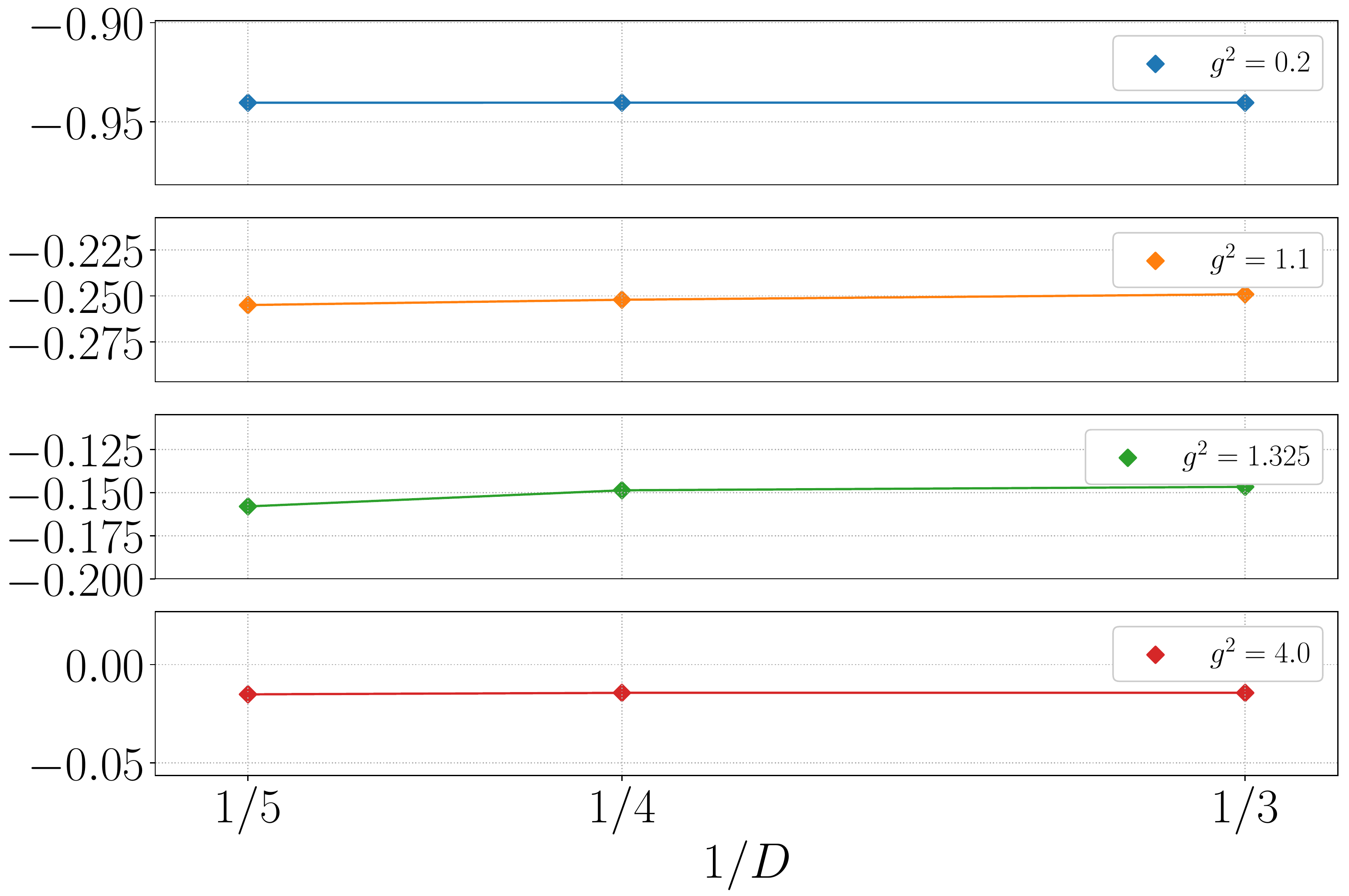}
\includegraphics[width=.49\columnwidth]{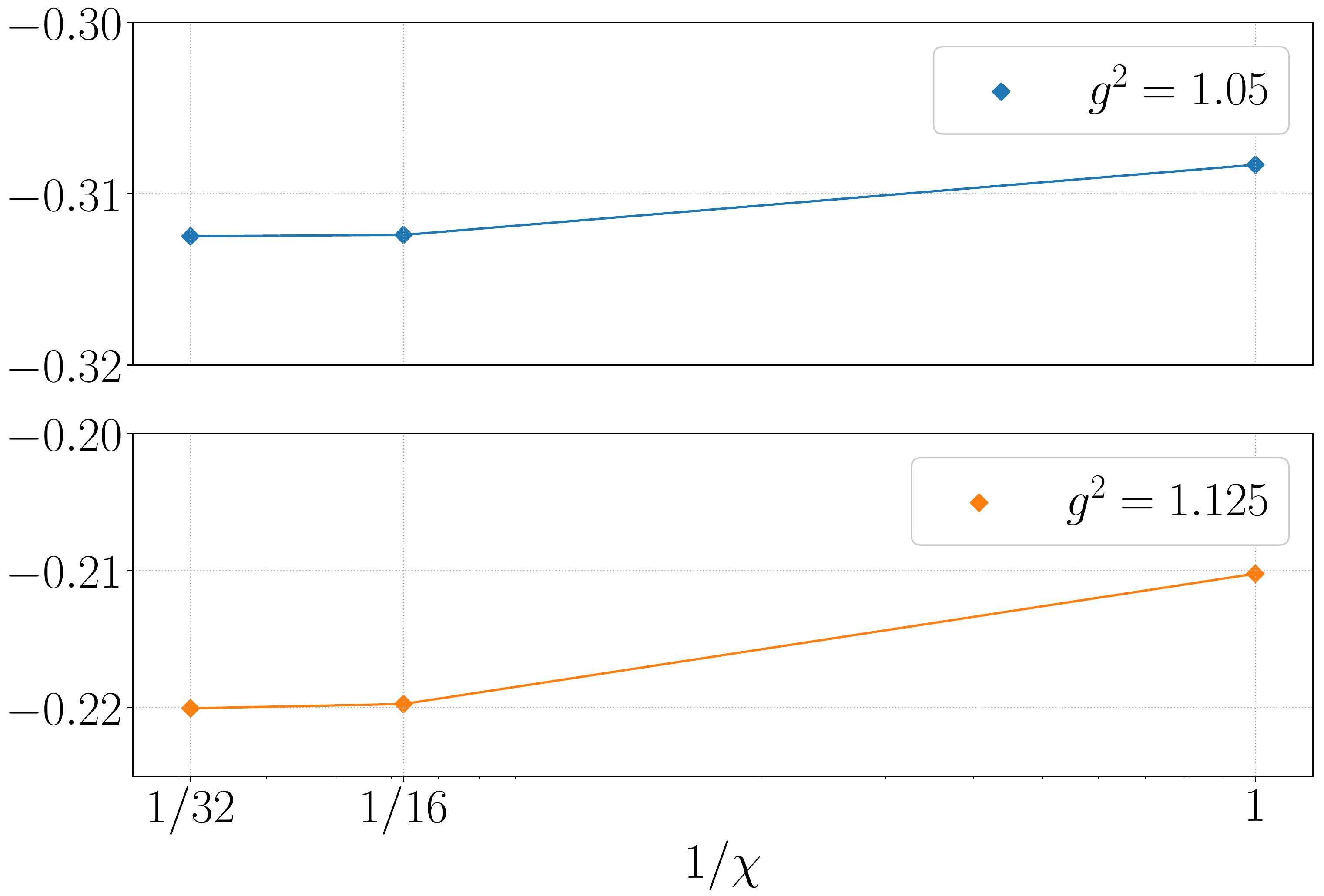}
\caption{Left: Ground state Energies $g^2 E_0$ for the zero charge sector for different
bond dimensions $(D=3,4,5)$. Right: Same quantity as a function of the number of states $\chi$ included in the CTM.}
\label{fig:errors}
\end{figure}

The left plot in \figref{fig:errors} shows for different values of the couplings our
results for the ground state energy for different values of the bond
dimension. It can be seen that at weak coupling the error is negligible. In fact,
the difference between $D=4$ and $D=5$ is less than $10^{-9}$. This is not
surprising, since the true ground state tends to a product state for $g^2 \to
0$. At intermediate couplings and near the phase transition the error rises up
to $6\%$ and stays rather constant up to strong couplings where the signal is so
weak that round-off errors start to become an issue.

When calculating expectation values via the CTM-method, it is crucial to ensure that the approximation of the environment has converged. To this end, it is customary to repeat the calculation of local observables 
with different number of states $\chi$
included in the environment. The right plot in \figref{fig:errors} shows for the case of $D=4$ the ground state energy from $\chi=1$ up to $\chi=32$. 

\begin{figure}
\includegraphics[width=.5\columnwidth]{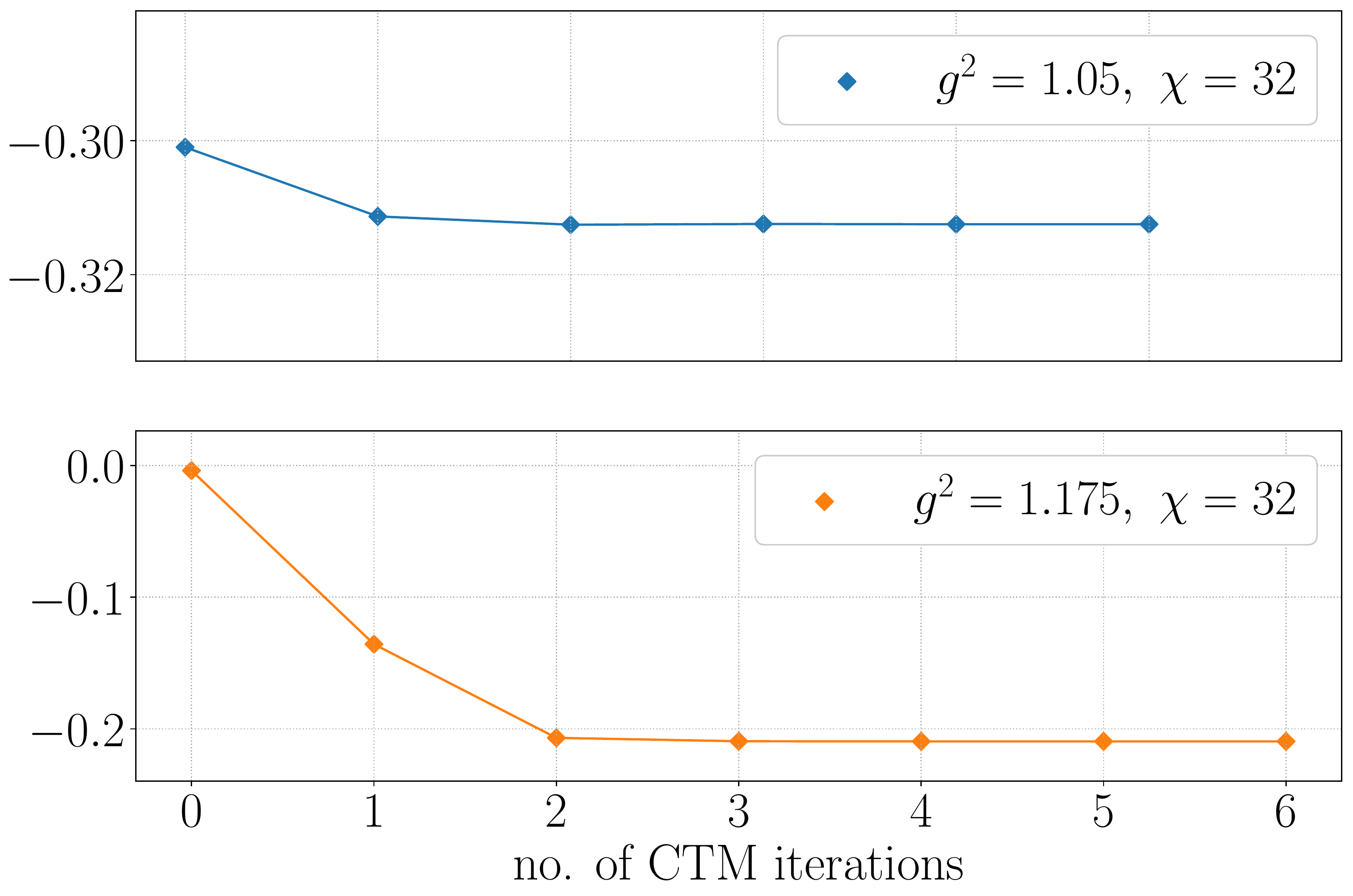}
\caption{Ground state Energies $g^2 E_0$ for $\chi = 32$ and $D=4$ as function of CTM-iterations.}
\label{fig:errors_ctm}
\end{figure}

Another convergence test can be performed at fixed value of $\chi$ by monitoring the change in expectation values as we keep absorbing unit-cells into the environment tensors. We have set $10^{-6}$ as the acceptable threshold for the difference in between iterations (see \figref{fig:errors_ctm}). From
these analysis we can conclude that the error is dominated by the bond dimension $D$ of the state.

\end{document}